\documentclass[12pt]{article}
\usepackage{geometry}
\usepackage{amssymb}
\usepackage{amsmath}
\numberwithin{equation}{subsection}
\usepackage{amsthm}
\usepackage{setspace}
\usepackage{caption}
\usepackage{cancel}
\usepackage{graphicx}
\DeclareGraphicsExtensions{.pdf,.png,.jpg}
\usepackage{relsize}
\usepackage{hyperref}
\usepackage{letltxmacro}
\usepackage[reflist=true,style=windycity]{biblatex}
\newtheorem{theorem}{Theorem}

\theoremstyle{definition}
\newtheorem{definition}{Definition}
\newtheorem{algorithm}{Algorithm}

\DeclareMathOperator*{\argmax}{arg\,max}

\begin{document}
\title{Characterizing nonatomic admissions markets}
\date{\today}
\author{Max Kapur\footnote{\emph{Email:} \href{mailto:maxkapur@snu.ac.kr}{maxkapur@snu.ac.kr}}}

\maketitle

\begin{abstract}
This article proposes a characterization of admissions markets that can predict the distribution of students at each school or college under both centralized and decentralized admissions paradigms. The characterization builds on recent research in stable assignment, which models students as a probability distribution over the set of ordinal preferences and scores. Although stable assignment mechanisms presuppose a centralized admissions process, I show that stable assignments coincide with equilibria of a decentralized, iterative market in which schools adjust their admissions standards in pursuit of a target class size. Moreover, deferred acceptance algorithms for stable assignment are a special case of a well-understood price dynamic called t\^{a}tonnement. The second half of the article turns to a parametric distribution of student types that enables explicit computation of the equilibrium and is invertible in the schools' preferability parameters. Applying this model to a public dataset produces an intuitive ranking of the popularity of American universities and a realistic estimate of each school's demand curve, and does so without imposing an equilibrium assumption or requiring the granular student information used in conventional logistic regressions.
\end{abstract}

\pagebreak
\tableofcontents

\pagebreak
\section{Introduction}
This article concerns admissions markets, which are dynamic systems consisting of students, schools, and a mechanism by which students are matched to schools each admissions cycle. An example of a mechanism is the application process: students apply to schools, schools admit their favorite students, and students choose their favorite school. Another kind of mechanism is a centralized admissions process in which the school board decides which students will attend which school. How do these mechanisms differ in their distributional outcomes, and how do these outcomes respond to a change in demographics, student preferences, or school quality?

The literature on admissions markets can be divided into one of two paradigms. First, there is a \emph{descriptive} paradigm, which uses observations of past market instances to predict the distribution of students across schools. For example, to achieve a target class size, college admissions offices want to know the probability that each applicant, if admitted, will choose to attend the school. This is an interesting problem because students vary in their endogenous preferences, in the number of competing programs to which they were admitted, and in their responsiveness to financial aid offers and marketing campaigns. A survey of sixteen colleges' admissions offices found that several had spent five- and six-figure sums to hire outside consulting firms to predict applicants' probability of enrollment \parencite[][]{estimatingapplications}. While the resultant models are proprietary, a classical technique in this area is logistic regression \parencite[][]{understandingandpredictingtheyield}.
 
On the demand side of the descriptive paradigm sits an international industry of admissions counselors who advise college applicants which universities to apply to and how to write their application materials. The admissions counselor plays an essential role in college admissions markets such as South Korea, where students are allowed to apply to only six colleges; to prevent a given applicant from wasting one of her applications, ruling out schools for which she is unqualified is imperative. Here, too, logistic regression and other classification methods are familiar instruments \parencite[][]{asimulationapproachtopredictingcollegeadmissions}. Given a comprehensive database of student profiles and school admissions standards, it is not difficult to imagine combining local models of demand curves and admissions standards to forecast the global allocation of students at every school. However, collecting the necessary data is prohibitively expensive. 

The second paradigm that has been widely applied to the study of admissions markets, more familiar to those in computational fields, is a \emph{prescriptive} paradigm that formulates the mapping of students to seats in schools as an optimization problem. In the ideal case, each student submits a preference order, or ranking, of the schools in the market, and each school likewise submits its preference order over the students. Then, the school board computes an assignment that maximizes, in some sense, the quality of the overall allocation. Literature on this so-called \emph{school-choice problem} has yielded a number of appealing assignment mechanisms, many of which stem from the classical deferred acceptance (DA) algorithm. DA's signature property is that it produces a stable matching, meaning no student feels cheated because a less qualified student has taken her seat at a school she favors. Variations of stable assignment include DA mechanisms that introduce randomness to break ties in schools' preference orders \parencite[][]{whatmatters}, and mechanisms that exploit the same ties to improve student utility \parencite[][]{expandingchoice} or optimize for distributional goals like gender parity \parencite[][]{distributionalgoals}. DA is itself the many-to-one form of the classical Gale--Shapley proposal algorithm for computing stable marriages, and its properties, including stability, incentive compatibility for the side of the market that takes the proposing role (usually students), and proposer optimality are well understood \parencite[][]{galeshapley1962, economicsofmatching}.

Mechanism designs such as DA presuppose a centralized assignment process in which students and schools abide by the school board's decision. Because DA is incentive compatible and produces fairly high (if suboptimal) welfare outcomes, this may not appear to be a hard sell. However, it is a matter of fact that many admissions markets, including most college admissions markets, are not centralized. Instead, each school sets its own admissions standards by consulting private goals regarding the number and kind of students it wishes to enroll. For example, in the United States, students are free to apply to as many colleges as time and their application-fee budget permits, and the fall admissions process concludes in April, when students observe which schools they were admitted to and commit to the school they like best. In this context, the notion of a school's ``capacity'' is fuzzier than the strict enrollment limit envisioned by DA, and represents instead the point at which the school believes that relaxing its admissions standards is not worth an additional student's tuition dollars. DA asserts that a centralized admissions process would be more ``efficient,'' but it does so by treating capacity as a hard constraint rather than as the optimum of each school's utility function. 

This article attempts to gain a broader view of admissions markets by characterizing them in a way that is agnostic to the assignment mechanism. The basis for my characterization is the nonatomic formulation of the school-choice problem, due to Azevedo and Leshno \parencite*{supplydemandfw}, which replaces individual students with a probability distribution over the space of possible preference lists and scores.\footnote{The nonatomic formulation of the school-choice has theoretical ancestors in the transportation science literature, where it is common to model traffic as a continuous flow rather than simulate the decisions of individual agents \parencite[][]{sometheoreticalaspectsofroadtraffic, theeconomicsofwelfare}. For this reason, I prefer the term \emph{nonatomic} to the other common term \emph{continuum}, which in the transportation literature refers to a mapping between nonatomic agents and the segment $[0,1]$ \parencite[as in][for example]{ksjthesis}} A nonatomic framing enables the characterization of stable matchings by a compact vector of school admissions cutoffs, which indicate the score threshold above which all students have the option of attending the school in question. As I argue, the set of distributional outcomes that can be characterized by cutoff vectors includes not only stable matchings, but also any assignment that can arise from a decentralized application process under the assumption that each school offers admission to the subset of applicants who exceed a certain rank in its preference list. This finding complements empirical research showing that stable matchings can predict outcomes in decentralized discrete marriage markets \parencite[][]{marryforwhat, matchingandsortinginonlinedating}.

A secondary goal of this article is to resolve the computational complexity inherent in previous descriptions of nonatomic admissions markets. A favorable property of the nonatomic formulation is that it can be interpreted as the limit of the discrete assignment problem as the number of students and seats increases to infinity; thus, score cutoffs in the nonatomic formulation are free of the ``noise'' associated with discretization. However, because the number of possible preference lists is factorial in the number of schools, and the set of possible consideration sets is exponential in the same, the space of possible student types is very large, and does not admit an obvious representation. While logistic regression models in the descriptive paradigm dodge this problem by taking students' personal characteristics as proximate indicators of their preferences, such models do not offer a clear picture of the interaction between students and schools, and deriving a parametric notion of school quality is difficult. Thus, the present study (especially its second half) seeks to harmonize the three goals of computational tractability, interoperability with publically available admissions data, and the ability to predict student distributions under both stable assignment and a decentralized application procedure.

\subsection{Organization}
The body of this article is divided into two sections. The first (\S\ref{interpeqinadmmkts}) establishes preliminary results concerning a certain notion of equilibrium in nonatomic admissions markets, whose several possible interpretations include stable assignment. While this section's foundation is a result of Azevedo and Leshno \parencite*{supplydemandfw} establishing the equivalence of stable matchings and Walrasian equilibrium in matching markets, I provide a straightforward proof. I also argue that even when we abandon the centralized school assignment mechanism implied by stable assignment algorithms in favor of a decentralized, iterative admissions market in which schools adjust their admissions standards in pursuit of a target class size, the same notion of equilibrium retains interpretive meaning. Moreover, I show that deferred acceptance algorithms are a special case of a price adjustment rule called t\^{a}tonnement, which is known to converge to equilibrium under certain conditions, and whose iterates behave like the price paths of perishable goods. 

In the second section (\S\ref{singlescoremodel}), I apply the results above to a parameterized admissions market that I call the single-score market with multinomial logit student preferences. This model is chosen for its computational tractability. Unlike the nonatomic markets theorized by previous work, in which computing the equilibrium requires evaluating a demand function that is exponentially complex in the number of schools, the model considered here admits a piecewise linear demand function, and each school's cutoff at equilibrium can be expressed as the solution of a triangular linear system. In this context, comparative statics at the equilibrium can be computed analytically. I also provide an inverse optimization procedure that computes the preferability parameter for each school given the cutoff and demand vectors, and I apply the procedure to a dataset of 677 American colleges. Despite the noisy input data and the model's intrinsic simplicity, this procedure yields a familiar ranking of top universities, and does so without using the costly opinion surveys or data on alumni outcomes that newspapers traditionally use to rank schools. Furthermore, it provides an estimate of each school's demand curve, which could be a useful supplement to the logistic regression models that program planners currently use to predict their admissions yield.

\section{Interpreting equilibria in admissions markets} \label{interpeqinadmmkts}
In this section, I define admissions markets and offer preliminary results on a notion of equilibrium. The applicability and interpretation of the equilibrium depends on the design of the admissions market. I offer three interpretations that reflect a variety of real-world market designs: It is a fixed point of a t\^{a}tonnement process in which schools adjust their cutoffs in pursuit of a target class size; it is a competitive equilibrium of a game in which schools' utility depends on the entering class size; and it is a stable matching. Next, I show that deferred acceptance algorithms can be viewed as t\^{a}tonnement processes in their own right, and sketch a t\^{a}tonnement algorithm for computing an equilibrium when an oracle is available for computing the demand.

\subsection{Admissions markets}
\begin{definition} An \emph{admissions market} consists of a set of schools $C = \{ 1\dots |C| \}$ and a mass-1 continuum of students over the set $S = C! \times [0, 1]^C$ of student types. The market is characterized by four parameters:
\begin{enumerate}
\item The measure $\eta: 2^S \mapsto [0, 1]$ over the continuum of students.
\item The score cutoff vector $p \in [0, 1]^C$. 
\item The demand vector $D \in [0, 1]^C$.
\item The capacity vector $q \in (0, \infty]^C$.
\end{enumerate}
\end{definition}

The model is \emph{nonatomic} in that it represents students as a probability measure over the set of student types instead of considering individual students as discrete actors. Each point $s$ in the set of student types $S$ is associated with a preference list over the schools $\succ_s$ and a percentile score at each school $\theta_{sc} \in [0,1]$. Hence, $S = C! \times [0, 1]^C$. 

Schools marginally prefer students with higher scores. Their admissions decisions are represented by the score cutoff vector $p$. Any student for whom $\theta_{sc} \geq p_c$ is said to be \emph{admitted} to school $c$. 

The demand vector represents the number of students who enroll at each school. Assume that each student attends her favorite school among the set of schools to which she is admitted, which is called her \emph{consideration set} $C^\# \in 2^C$. Then the demand for school $c$ is a function of $p$ and $\eta$; specifically, it is the measure of students who are admitted to school $c$ and not admitted to any school that they prefer to $c$:
\begin{equation} \label{demanddefinition}
D_c \equiv \eta\left(s: c = \argmax_{\text{wrt } \succ_s} \left\{\hat c: \theta_{s\hat c} \geq p_{\hat c} \right\}\right)
\end{equation}
Observe that $D_c$ is weakly decreasing in $p_c$ and weakly increasing in $p_{c'}$ for $c' \neq c$, and that $p_c = 1 \implies D_c = 0$ regardless of the other schools' cutoffs.

If the preference lists are independent of the score vectors, then the demand can also be expressed as the sum of the demand from each combination of preference list and consideration set:
\begin{gather} \label{demandbigsum}
\begin{aligned}
D_c = 
\sum_{\succ_s \in C!} \sum_{\substack{C^\# \in 2^{C}:\\ c \in C^\#}}
\eta\Big(\;s:&~~\underbrace{\theta_{sc'} \geq p_{c'}, \forall c' \in C^\#}_{\text{got into schools in } C^\# } \\
\text{ and} &~~\underbrace{\theta_{sc''} < p_{c''}, \forall c'' \in C \setminus C^\#}_{\text{rejected elsewhere}} \\
\text{ and} &~~\underbrace{c \succ_s \hat c, \forall \hat c \in C^\#\setminus \{c\}}_{\text{prefers } c \text{ among } C^\#} \;\Big)
\end{aligned}
\end{gather}
This expression yields immediate insight into the complexity of nonatomic admissions markets. When schools are allowed to set their own cutoffs, and students are given free choice among the schools in their consideration sets, the number of terms in the sum above is $|C|!\times2^{|C|}$. Fortunately, such a general characterization of students is not always needed. The model considered in the second section of this paper (\S\ref{singlescoremodel}) characterizes $S$ using only a $|C|$-vector of school preferability parameters. Alternatively, generic markets can be discretized by sampling individual students' preference lists and score vectors. 

Each school's \emph{capacity} $q_c > 0$ represents the fraction of the total mass of students that the school can accept. The capacity is used to define the notion of equilibrium below. 

This article assumes that students prefer to be assigned to any school, even their last choice, than to remain unassigned. This is without loss of generality: If some students prefer to be unassigned than attend a particular school, then this choice can be incorporated into the model by adding a dummy school, representing nonassignment, with arbitrary large capacity.

Also, assume that almost no ties occur among the scores at a given school. That is, for any fixed $c$ and constant $\bar \theta$, $\eta( s: \theta_{sc} = \bar \theta) = 0$. Then, by transforming the scores at each school by the inverse of its marginal cumulative distribution function, we may assume without loss of generality that for a random student $s$, $\theta_{sc} \sim \operatorname{Uniform}(0, 1)$. It follows that the demand function given in Equation \eqref{demandbigsum} is continuous in $p$. 

\subsection{Notion of equilibrium}
The notion of equilibrium defined below applies to an admissions market in which each school has a fixed capacity $q_c$. Prescriptivist solutions to school-choice problems, such as the deferred acceptance algorithm, interpret this capacity as a hard constraint on the number of students each school is allowed to accept. Such hard constraints are possible in reality, to the extent that physical space restrictions or government regulation constrain the number of students each school can accept. However, as argued in the following section, we can also regard the capacity as a \emph{target} number of students that represents the school's estimate of the class size at which an additional student's tuition dollars are not worth the relaxation of admissions standards required to recruit her.

\begin{definition} \label{marketeqconditions} An admissions market is in \emph{equilibrium} if the following conditions hold:
\begin{align} D_c &\leq q_c, \quad \forall c \label{capacitycondition} \\
D_c &= q_c, \quad \forall c: p_c > 0 \label{stabilitycondition}
\end{align}
The first condition, called the \emph{capacity condition,} says that no school's demand exceeds its capacity. The second, called the \emph{stability condition,} says that if a school is rejecting students, it must be at full capacity.
\end{definition}

Using the sign constraint on $p$ and the capacity condition, the stability condition may be rewritten as $D_c < q_c \implies p_c = 0$ or as $p^T \left(D - q\right) = 0$.

As shown below, a sufficient condition for the existence of the equilibrium is that the demand is continuous in $p$. A sufficient condition for the uniqueness of the equilibrium is that the demand is strictly decreasing in $p$. Both of these follow from, for example, an assumption of full support in $\eta$. 

Before interpreting these conditions, it is worth stating a helpful fact. 

\begin{theorem}If $\hat p$ satisfies the equilibrium conditions, then it is a market-clearing cutoff vector. That is, the total measure of assigned students is $\min\{1, \sum_c q_c\}$
\end{theorem}

\begin{proof}Let $\hat \eta$ denote the measure of assigned students: $\hat \eta \equiv \sum_c D_c(\hat p)$. By the capacity criterion, we have $\hat \eta \leq \sum_c q_c$. Since every student prefers assignment to nonassignment, $\hat \eta = 1 - \eta ( s: \theta_{sc} < p_c, \forall c \in C) \leq 1$. If at least one school has $p_c = 0$, then $\hat \eta = 1 \leq q_c$ and the statement holds. Otherwise, the stability condition applies to every school, meaning $\hat \eta = \sum_c q_c \leq 1$. \end{proof}

In light of this observation, Azevedo and Leshno \parencite*{supplydemandfw} call the equilibrium conditions the market-clearing conditions. I will avoid this terminology, because it is possible for a centralized mechanism to clear the market without using score cutoffs at all (for example, by assigning students to schools randomly). 

\subsection{Interpretation of the equilibrium conditions}
I offer three interpretations of the equilibrium conditions. With additional assumptions on $\eta$, such as full support, all three interpretations become sufficient conditions for equilibria as well. Additionally, the first interpretation establishes a sufficient condition for the existence of an equilibrium. In all cases, assume the distribution of student types $\eta$ and the school capacities $q$ are fixed; thus, the demand vector $D(p)$ is determined entirely by the cutoffs $p$. 

\subsubsection{As a fixed point of a t\^{a}tonnement process} \label{asafixedpoint}
Suppose that each school has set an admissions target of $q_c$ and observes its demand from year to year. If more than $q_c$ students enroll, then the school attempts to reduce remand by increasing its cutoff. If fewer than $q_c$ students enroll, the school attempts to increase demand by lowering its cutoff (but not past zero). 

Let $Z(p) \equiv D(p) - q$ denote the excess demand vector. Then the process described above implies the following recursive relation between the cutoff vector in year $k$ and in year $k+1$.
\begin{align} \label{tatonnementdef}
p_c^{(k+1)} = \max\biggl\{0,~p_c^{(k)}+ \Gamma_c\Bigl[Z_c\bigl(p^{(k)}\bigr)\Bigr] \biggr\}
\end{align}
Here $\Gamma$ is any sign-preserving function. Such a dynamic process, in which prices adjust in the direction of excess demand, is called a \emph{t\^{a}tonnement process}.
\begin{theorem}
If $\bar p$ is a fixed point of the t\^{a}tonnement process, then it satisfies the equilibrium conditions. The converse also holds.
\end{theorem}
\begin{proof} Pick $\bar p$ such that $\bar p = \max\bigl\{0, \bar p_c + \Gamma_c\left[Z_c(\bar p)\right] \bigr\}$. Subtract $\bar p_c$ from both sides to obtain $0 = \max\bigl\{-\bar p_c,\Gamma_c\left[Z_c(\bar p)\right] \bigr\}$, which implies $0 \geq \Gamma_c\left[Z_c(\bar p)\right]$. This means that the excess demand is nonpositive, which establishes the capacity condition. Now, suppose $\bar p_c > 0$; then $\bar p_c =  \bar p_c + \Gamma_c\left[Z_c(\bar p)\right] $ establishes the stability condition $Z_c(\bar p = 0)$. Hence, any fixed point of the t\^{a}tonnement process is an equilibrium.\footnote{This conception of the equilibrium conditions as a ``fixed point'' is similar in name but distinct from a known result in lattice theory by Echenique and Oviedo \parencite*{coremanytoonebyfixedpoint}, which applies to discrete many-to-one matchings.}

As for the converse, if $p^{(k)}$ satisfies the equilibrium conditions, it is easy to see that $p^{(k+1)} = p^{(k)}$. \end{proof}

Moreover, if the demand function is continuous in $p$, then Brouwer's fixed-point theorem guarantees that a fixed point exists, because the cutoff update maps the compact convex set $[0, 1]^C$ to itself. This means that continuous demand is sufficient for the existence of an equilibrium in admissions markets. 

Quantitative economists have studied t\^{a}tonnement processes extensively. For an introduction, see Codenotti and Varadarajan \parencite*{compmkteq} or Intriligator \parencite*[][chap. 9]{mathematicaloptandecontheory}. A classical proof of various convergence conditions is Uzawa \parencite*{walrastatonnement}. 

\subsubsection{As a competitive (Nash) equilibrium} \label{asacompeq}
\begin{figure}
\begin{center}\includegraphics[width=0.7\linewidth, ]{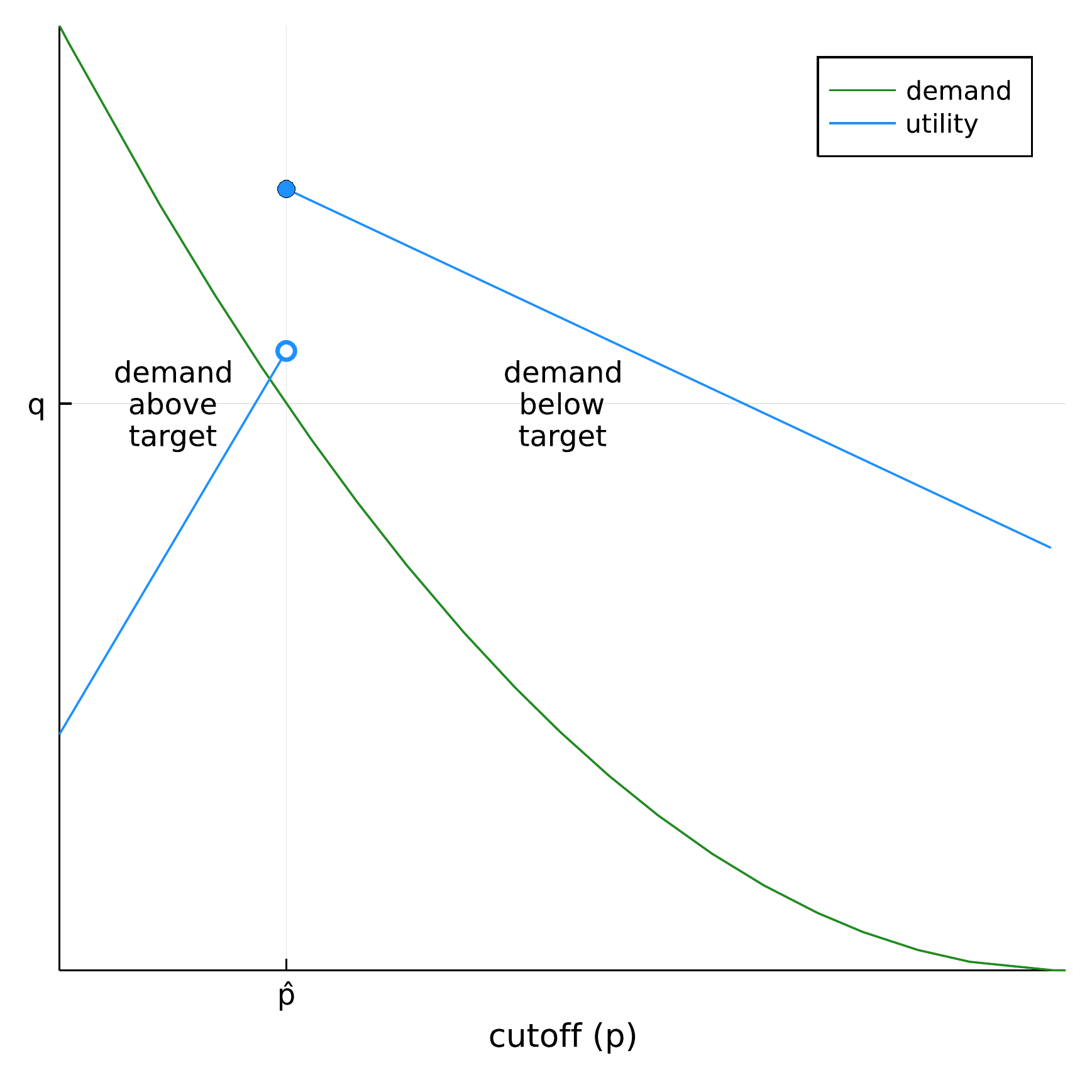}\end{center}
\captionsetup{singlelinecheck=off}
    \caption[.]{A stylized illustration of conditions under which stable assignment is a competitive equilibrium. The school's utility function is maximized when its demand equals the target demand $q$. The cutoff value $\hat p$ that attains this maximum may depend on the cutoffs at the other schools.}
\label{stylized-utility}
\end{figure}

Suppose that each school's capacity $q_c$ is a physical constraint on the number of students it can admit. Each school would like to recruit as many students as possible. However, if more students choose to attend the school than the school has capacity for, it must rent additional classroom space at considerable expense. Pick a school $c$ and fix the cutoffs at the other schools $p_{c'}$ . Let $u_c(p_c; p_{c'})$ denote school $c$'s utility function, and let $\hat p_c \equiv p_c: D_c(p_c; p_{c'}) = q_c$ denote the cutoff value that causes $c$ to fill its capacity, if such a value exists. In the situation described, $u_c$ is increasing in $p_c$ when $0 \leq p_c < \hat p_c$, decreasing in $p_c$ when $p_c \geq \hat p_c$, and the fixed costs associated with excess demand are nonnegative: \[\lim_{p_c \to \hat p_c^-} u_c(p_c; p_{c'}) \geq u_c(\hat p_c; p_{c'})\]
Hence, $\hat p_c$ maximizes the school's utility. A stylized illustration of a utility function and demand curve meeting this condition appears in Figure \ref{stylized-utility}. 

Another scenario in which utility functions with this shape may arise is as follows: Schools' utility functions are determined mostly by the number of students they enroll, and to a lesser extent by their average score. If the demand for a school is less than its capacity, then a marginal student is always desirable. On the other hand, if the demand exceeds the capacity, then the school has no ability to procure space for the excess demand. Instead, it allows students to register on a first come, first served basis. Because the set of registered students is a random subset of the students who attempt to enroll at the school, it provides the school with less overall utility than if it handpicked the $q_c$ students with the highest scores. That is, $u_c$ is increasing when $0 \leq p_c < \hat p_c$, and decreasing when $p_c \geq \hat p_c$. 

In both of these situations, the admissions market equilibrium can be interpreted as a Nash equilibrium. 
\begin{theorem}
Consider the game in which each school picks a cutoff $p_c \in [0,1]$ and tries to maximize a utility function $u_c(p_c; p_{c'})$ whose optimum in $p_c$ occurs when $D_c(\hat p_c; p_{c'}) = q_c$ for some constant $q_c$, and which is decreasing in $p_c$ when $D_c \leq q_c$. Call $\bar p$ a Nash equilibrium of this game if, for each school $c$, the following holds.
\[\bar p_c \in \argmax_{\pi_c} \left\{ u_c(\pi_c; p_{c'}) : p_c \in [0,1] \right\}\]
If $p^*$ satisfies the equilibrium conditions, then it is a Nash equilibrium of the game above. If the utility functions are strictly increasing or decreasing, the converse is also true.
\end{theorem}
\begin{proof} Suppose $p^*$ satisfies the admissions market equilibrium conditions. For the schools for which $D_c = q_c$, their utility is globally maximal. For a school for which $D_c < q_c$, the only way for the school to increase its utility is to decrease its cutoff, but by assumption, $p_c^* = 0$. Hence, is incentivized to change its cutoff, and $p^*$ is a Nash equilibrium of the game defined by the schools' utility functions and the action space $p_c \in [0, 1]$.

The converse can be shown similarly.\end{proof}

Both of the above are natural situations in which the t\^{atonnement} dynamics may arise.

\subsubsection{As a stable matching}
The final interpretation of the equilibrium conditions comes from Lemma 1 of Azevedo and Leshno \parencite*{supplydemandfw}, which says that there is a one-to-one relationship between stable matchings and equilibrium cutoff vectors. Here, I define stable matchings and offer a proof of their lemma. 

The notion of a stable assignment has its roots in the field of mechanism design, and thus emerges most naturally from a centralized school choice process as follows: Before the school year begins, students submit to the school board a form indicating their preference order over the set of schools in the district. Likewise, schools indicate their preference order over the students (or equivalently, the scores they have given to each student). Then, the school board determines the assignment of students to schools.

First, some notation. A assignment is a mapping of students to schools.
\begin{definition}
A school choice \emph{assignment} is a mapping $\mu: S \to C \cup \{c_0\}$. $\mu(s) = c_0$ represents nonassignment.
\end{definition}
The school board is interested in matchings, or assignments that respect capacity constraints. 
\begin{definition}
A \emph{matching} is an assignment $\mu$ that respects schools' capacity constraints; namely, $\eta (s: \mu(s) = c) \leq q_c, \forall c \in C$. 
\end{definition}
To protect itself from lawsuits and encourage honest participation in the assignment process, the school board decides to rule out matchings that create ``justified envy'' \parencite[][7]{expandingchoice}---that is, matchings in which a student $s$ who prefers school $c$ to $c'$ is assigned to $c'$ despite scoring higher than a student assigned to $c$, or $c$ having remaining capacity. In such a situation, $(s, c)$ is called a blocking pair, and a stable matching is a matching that does not admit any blocking pairs.
\begin{definition}
A \emph{stable matching} (or \emph{stable assignment}) is a matching $\mu$ that admits no blocking pairs. That is, there exist no student--school pairs $(s, c)$ such that $c \succ_s \mu(s)$ and one of the following holds:
\begin{itemize}
\item School $c$ has remaining capacity:
\[\eta\bigl(s: \mu(s) = c \bigr) < q_c\]
\item (And/or,) school $c$ has admitted an inferior student:
\[\exists s' \neq s: \mu(s') = c \text{ and } \theta_{s'c} < \theta_{sc}\] 
\end{itemize}
These are called type I and type II blocking pairs, respectively.
\end{definition}

In a nonatomic admissions market, where the number of students is infinite, the definition above offers no guidance as to how to encode a stable matching $\mu$. However, it turns out that stable matchings are in one-to-one correspondence with equilibrium cutoff vectors. Therefore, any stable matching $\mu$ can be fully encoded by a $|C|$-vector of cutoffs $p$. This fact is invaluable in a computional context.

To establish this result, we must define operators that take cutoff vectors to assignments, and vice-versa. First, the assignment of students induced by instating the cutoff vector $p$ and allowing students to choose freely among their consideration set is
\[\mu_p(s) \equiv \max_{\succ_s} c \in C : \theta_{sc} \geq p_c, \qquad \forall s \in S\] 
Note that this assignment is not necessarily a matching because it may violate the capacity constraints. 

Second, the following expression gives the admissions cutoffs implied by a given matching $\mu$, namely, the minimum score of the students admitted to each school:
\[p_c(\mu) \equiv \min \left\{\theta_{sc}: \mu(s) = c\right\}\]
Both operators were considered by Azevedo and Leshno \parencite*{supplydemandfw}. In general, these operators are not necessarily inverses of each other. However, as implied by the following theorem, they are inverses when we restrict their domains to the sets of equilibrium cutoffs and stable matchings, respectively. 

\begin{theorem}If $p^*$ satisfies the equilibrium conditions, then $\mu_{p^*}$ is a stable matching. 

Conversely, if $\bar \mu$ is a stable matching and $\eta$ has full support, then $p_c(\bar \mu)$ satisfies the equilibrium conditions. \end{theorem}

\begin{proof}Pick an equilibrium cutoff $p^*$. Then by the definitions of the demand function (equation \eqref{demanddefinition}) and equilibrium conditions (equation \eqref{marketeqconditions}), 
\begin{align*}
D_c(p^*) = \eta\bigl(s: \mu_{p^*}(s) = c\bigr) &\leq q_c, \quad\forall c \in C \\
D_c(p^*) = \eta\bigl(s: \mu_{p^*}(s) = c\bigr) &= q_c, \quad \forall c: p_c^* > 0 
\end{align*}
By the capacity condition, no school exceeds its capacity, so $\mu$ is an assignment. By the stability condition, there are no type I blocking pairs. And there are no type II blocking pairs, because if a student fails to meet the cutoff for a school she prefers to $\mu_{p^*}(s)$, it is because the school has replaced her with students who got higher scores. Hence, $\mu_{p^*}$ is a stable matching.

The converse is proven as follows. Fix a stable matching $\bar \mu$, and let $\bar p \equiv p(\bar \mu)$. To get a contradiction, suppose $\bar p$ is not an equilibrium. This can happen in two ways:
\begin{itemize}
\item For some school $c$, $D_c( \bar p) > q_c$. This means $D_c( \bar p)  >  \eta\bigl(s: \bar \mu(s) = c\bigr) $, which implies the existence of a student $s$ who is admitted to $c$ at $\bar p$ (that is, $\theta_{sc} \geq \theta_{s'c}$ for some $s': \bar \mu(s') = c$) and prefers $c$ among her consideration set, but for whom $\bar \mu(s) \neq c$. Then $(s, c)$ is a type II blocking pair; hence, $\bar \mu$ is not a stable matching.
\item For some school $c$, $\bar p_c > 0$ and $D_c(\bar p) < q_c$. By full support, there is a student $s$ for whom $c \succ_s \bar \mu(s)$ and $\theta_{sc} < \bar p_c$. The latter implies that $\bar \mu(s) \neq c$. Hence, $(s, c)$ is a type I blocking pair, and $\mu$ is not a stable matching. 
\end{itemize}
Therefore, $\bar p$ must satisfy the equilibrium conditions.
\end{proof}

\subsection{Deferred acceptance algorithms as t\^{atonnement} processes} \label{defaccaretat}
The classical solution to the stable matching problem is known as the deferred acceptance (DA) mechanism, which comes in many flavors. When neither students' nor schools' preference lists contains ties, the student-proposing DA procedure is a deterministic algorithm for a stable matching.  In this section, I first define the student- and school-proposing DA algorithms. Then, I show that DA algorithms are t\^{a}tonnement processes.

\begin{algorithm} \label{studentproposingDA}
The \emph{student-proposing deferred acceptance algorithm} is as follows. Given each student's preference order $\succ_s$ over the set of schools, without ties;  each school's score distribution $\theta_{.c}$ over the set of students, with zero probability of ties; and the capacity $q_c$ of each school, the following steps are repeated until no rejections take place:
\begin{enumerate}
\item Each student applies to the school highest on her list.
\item Each school examines the applications it received. If it received more applicants than it can seat, it rejects its least-favorite applicants such that the remaining applicants fill its capacity exactly.
\item Each rejected student removes the school that rejected her from her list.
\end{enumerate}
When the algorithm terminates, return the assignment $\mu$, where $\mu(s)$ is highest school remaining on $s$'s preference list, or $c_0$ if no schools remain. 
\end{algorithm}
The properties of the resultant assignment are well known: In the discrete case with $|S|$ students and $|C|$ schools, the algorithm terminates in at most $|S||C|$ iterations. The resultant matching $\mu$ is stable. $\mu$ is also strongly student optimal, meaning that if another assignment $\mu'$ is chosen from the set of stable matchings, then $\mu(s) \succeq_s \mu'(s)$ for all students $s$, and there is at least one student for whom $\mu(s) \succ_s \mu'(s)$. Finally, the algorithm is weakly incentive compatible for individual students. That is, no student can obtain a better match than $\mu(s)$ by falsifying her preference list. Succinct proofs of these results are given in Roth \parencite*{economicsofmatching}. The DA algorithm was first proposed by Gale and Shapley \parencite*{galeshapley1962}, and DA was applied to school choice by Abdulkadiro\u{g}lu and Tayfun S\"{o}nmez  \parencite*{schoolchoiceamechanismdesignapproach}. 

It is worthwhile to compare school-proposing DA. 
\begin{algorithm}
The \emph{school-proposing deferred acceptance algorithm} is as follows. Given each student's preference list $\succ_s$ over the set of schools and each school's scores $\theta_c$ over the set of students, both without ties, and the capacity $q_c$ of each school, the following steps are repeated until no rejections take place:
\begin{enumerate}
\item Each school proposes to the $q_c$ applicants in its consideration pool. If fewer than $q_c$ students are left, the school proposes to all remaining students.
\item Each student examines the proposals she received, rejecting all but her favorite.
\item Each school removes students who rejected it from its consideration pool.
\end{enumerate}
When the algorithm terminates, return the assignment $\mu$, where $\mu(s)$ is the school $s$ prefers among those that proposed to her, or $c_0$ if she received no proposals.
\end{algorithm}
When each school has room for only one student (that is, in the marriage problem), this algorithm has symmetrical properties to those of forward DA, including student pessimality and incentive compatibility for the schools.\footnote{These results do not extend to the general case, because any notion of school optimality and school incentive compatibility requires knowledge of schools' preference lists over \emph{sets} of students, rather than individuals \parencite[][]{collegeadmissionsisnotmarriage}.} For these reasons, student-proposing deferred acceptance is seldom used in school choice, and its counterpart in the National Residency Matching Program was abandoned in favor of a resident-proposing algorithm. However, in practice the differences among the resulting assignments tend to be minor \parencite[][]{unbalancedrandommatchingmarkets}. 

\begin{theorem}When $\eta$ has full support, the student-proposing DA algorithm is a t\^{a}tonnement process in which the initial cutoff vector is $p = \vec 0$, and the school-proposing DA algorithm is a t\^{a}tonnement process in which the initial cutoff vector is $p = \vec 1$. \end{theorem}
\begin{proof}
Consider the case of student-proposing DA, and let $\mu^{(k)}$ denote the tentative assignment formed at each iteration of the algorithm. That is, $\mu^{(k)}(s)$ is the school at the top of $s$’s preference list at the beginning of the $k$th iteration.

It suffices to prove the following three statements.
\begin{enumerate}
\item Each iterate $\mu^{(k)}$ is characterized by a cutoff vector $p^{(k)}$.
\item The initial cutoff vector has $p^{(0)} = \vec 0$.
\item $p^{(k+1)}$ is related to $p^{(k)}$ by a t\^{a}tonnement update.
\end{enumerate}
The statements are shown as follows. 
\begin{enumerate}
\item Fix $k$, let $p^{(k)} \equiv p\left(\mu^{(k)}\right)$, and let $m = \mu_{p^{(k)}}$. I will show that $m = \mu^{(k)}$. Pick a student $s$ and let $c = m(s)$. Since $s$ is among the set of students who determined the cutoff vector $p^{(k)}$, $ \mu^{(k)}(s)$ is still in $s$’s consideration set under these cutoffs; hence,  $c \succeq_s \mu^{(k)}(s)$.  Now suppose $c \succ_s \mu^{(k)}(s)$. Since $s$ prefers $c$, she must have been applied to $s$ in a previous round and been rejected. This implies $p^{(k)}_{c} > \theta_{sc}$; hence $s$ is not admitted to $c$ in $m$, a contradiction. It follows that $m(s) = \mu^{(k)}(s)$. 

\item At the beginning of the first iteration, each student is tentatively assigned to her favorite school. By the support assumption, for all schools $c$, the set of students who have $c$ at the top of their list and whose score is almost zero is nonempty. Hence, the minimum score over the tentative assignment at each school is zero.

\item At the beginning of each iteration of student-proposing DA, students who were not rejected in the previous iteration apply to the same school as before; on the other hand, students who were rejected apply to new schools. This means that students who apply to school $c$ at the $k$th iteration and are \emph{not} rejected are a subset of the set of students who apply at the $k+1$th iteration; hence, at every school $c$, $p_c^{(k+1)} \geq p^{(k)}$.

Suppose $p_c^{(k+1)} > p_c^{(k)}$. This implies that $c$ rejected students during iteration $k$, which is true only if the number of students tentatively matched to $c$ at $k$ exceeded $c$’s capacity. Hence, $D_c ( p^{(k)}) > q_c$, and the excess demand $Z_c(p^{(k)})$ was positive at $k$. This agrees with the sign of the change in cutoff. 

Suppose $p_c^{(k+1)} = p^{(k)}$. This means that $c$ made no rejections during the $k$th iteration, or equivalently, that the number of students tentatively assigned to $c$ is less than or equal to $q_c$. In our notation, $\eta(s: \mu^{(k)} (s) = c)  = D_c (p^{(k)}) \leq q_c$. Hence the excess demand $Z_c(p^{(k)})$ is nonpositive. If $Z_c(p^{(k)}) = 0$, then the statement holds. If $Z_c(p^{(k)}) < 0$, then the statement holds only if $p_c^{(k)} = 0$. To get a contradiction, suppose $p_c^{(k)} > 0$. By the support assumption, there are students whose score is less than $p_c^{(k)}$ who have ranked $c$ first. These students applied to $c$ in an earlier round---call it $j$---and were rejected. This implies $c$ filled its capacity at $j$. Since the students not rejected at $j$ continue to apply to $c$ unless rejected again, $c$ fills its capacity at all subsequent rounds, including round $k$. Hence $Z_c(p^{(k)}) \geq 0$, a contradiction.
\end{enumerate}
The case of school-proposing DA is analogous. 
\end{proof}

Using this result, we can rewrite the DA algorithms above in a ``computational'' form that uses the cutoff vector $p$ as the state variable. In fact, allowing $p^{(0)}$ to take an arbitrary value, we can define a whole subclass of t\^{a}tonnement processes that use deferred acceptance to update the cutoff vector. I conjecture that this process converges regardless of the value of the initial cutoff vector. However, even if that conjecture is true, we still have some distance to tread before arriving at a general algorithm for admissions market equilibrium, because the process defined below does not specify how to compute the demand vector or its roots.
\begin{algorithm}
A \emph{deferred acceptance t\^{a}tonnement process} is as follows. Given an initial cutoff vector $p^{(0)}$, each student's preference order $\succ_s$ over the set of schools, without ties;  each school's score distribution $\theta_{.c}$ over the set of students, with zero probability of ties; and the capacity $q_c$ of each school, the following steps are repeated until $p^{(k+1)} = p^{(k)}$: For $k = 0, 1, \dots$, 
\begin{enumerate}
\item Compute the demand vector $D (p^{(k)})$. 
\item For each school $c$ for which $D_c > q_c$, increase the cutoff so that
\[p_c^{(k+1)} \equiv p_c: D_c(p_c; p_{c'} ) = q_c\]
\item For each school $c$ for which $D_c < q_c$, decrease the cutoff (but not past zero) so that
\[p_c^{(k+1)} \equiv \begin{cases}
p_c: D_c(p_c; p_{c'} ) = q_c, &\text{if such a } p_c \text{ exists}\\
0, &\text{otherwise}
\end{cases}\]
\item Otherwise, let $p_c^{(k+1)} \equiv p_c^{(k)}$.
\end{enumerate}
When the algorithm terminates, return $p_c^{(k)}$. 
\end{algorithm}

This algorithm bears a strong resemblance to the so-called successive t\^{a}tonnement process in which each company adjusts its price to the value that clears its supply under the assumption that other companies' prices are fixed \parencite[see][equation 6]{walrastatonnement}. In the game-theoretic interpretation, the cutoff update can be interpreted as each school playing its ``best response'' to the strategies demonstrated by its opponents \parencite[][\S19.3.4]{networkformationgames}. 

\subsection{Computing the equilibrium} \label{computingtheeq}
With the results above in hand, consider a general admissions market in which $\eta$ and $q$ are fixed. We want to compute the equilibrium of this market. It is impractical to apply a DA algorithm to nonatomic admissions markets, because DA requires using exact line search to determine the new cutoff value for each school, and in general the demand is difficult to compute.

A moderate improvement over the deferred acceptance t\^{a}tonnement process is to use a \emph{simultaneous} t\^{a}tonnement process \parencite[][equation 3]{walrastatonnement} that evaluates the demand vector once per iteration and updates the cutoffs in the direction of the excess demand according to a predetermined sequence of decreasing step sizes. Under a light assumption on $D$, such as continuity, this process can be used to compute the equilibrium to arbitrary precision.
\begin{algorithm} \label{admissionseqtatalgo}
The \emph{admissions equilibrium t\^{a}tonnement algorithm} is as follows. Given an initial cutoff vector $p^{(0)}$, market parameters $\gamma$ and $q$, step parameters $\alpha >0$ and $0 \leq \beta < 1$, and a tolerance parameter $\epsilon$:
\begin{enumerate}
\item Compute the excess demand $Z = D (p^{(k)}) - q$. 
\item Update the cutoffs:
\[ p_c^{(k+1)} \equiv p_c^{(k)} + \frac{\alpha}{(k+1)^\beta} Z_c\]
\item Terminate if $| p_c^{(k+1)} - p_c^{(k)} | < \epsilon, \forall c$; otherwise, set $k \equiv k+1$ and repeat. 
\end{enumerate}
When the algorithm terminates, return $p_c^{(k)}$. 
\end{algorithm}
If the demand is continuous, convergence to an $\epsilon$-approximate equilibrium is guaranteed by the fact that the sequence of step sizes satisfies the Robbins--Monro conditions \parencite[][]{robbinsmonro}. However, the algorithm is not necessarily computationally efficient. Although a good choice of parameters can enable the algorithm to terminate in a small number of iterations, in general, evaluating the demand vector at each iteration incurs a high computational cost. For example, even under the assumption of independence between students' preference orders and score vectors, the number of terms in each school's demand can be $|C|! \times 2^{|C|}$, as shown in equation \eqref{demandbigsum}.

Alternatively, if we have the ability to sample student preference lists and score vectors from $\eta$, then we can exploit the relationship between equilibrium cutoffs and stable matchings to estimate the equilibrium cutoffs with high confidence in polynomial time. The technique is as follows: Draw a discrete sample from $\eta$ and run a DA algorithm. Then compute the minimum score at each school in the resultant stable assignment. If student-proposing DA is used, the expected value of each the cutoffs from the student-optimal stable match approaches the equilibrium cutoff value from below as the size of the sample goes to infinity \parencite[][]{supplydemandfw}. Similarly, if school-proposing DA is used, the expected value of the obtained cutoffs approaches the equilibrium from above. 

In summary, we have an in t\^{a}tonnement an expensive, guaranteed-precision technique, and in discrete DA a cheap, stochastic technique for computing the equilibrium. Neither is completely satisfactory. One motivation for the model considered in the second half of this article (\S\ref{singlescoremodel}) is the fact that the equilibrium can computed in closed form by solving a linear system in $|C|$ equations for $p$.

\subsection{Equivalent formulations of the equilibrium conditions}
The conditions for a market-clearing cutoff vector given in definition \ref{marketeqconditions} can be expressed in a few additional ways. Throughout this section, assume $\eta$ and $q$ are fixed and use \[F(p) \equiv -Z(p) = q - D(p)\]
to denote the excess supply vector at $p$. 

\subsubsection{Nonlinear complementarity problem}
By inspection, the market-clearing cutoff problem is equivalent to the following nonlinear complementarity problem:
\begin{gather} \label{nonlinearcompprob}
\begin{aligned}
\text{find } p:\quad F(p)^T p & = 0 \\ F(p) &\geq 0 \\ p & \geq 0
\end{aligned}
\end{gather}

\subsubsection{Variational inequality problem}
By a canonical result, the following variational inequality problem is also equivalent:
\begin{align*}
\text{find } p \geq 0:\quad F(p)^T (\pi-p) \geq 0, \quad \forall \pi \geq 0
\end{align*}
If $D_c$ is strictly decreasing in $p_c$, then $F_c$ is strictly increasing, and $p^*$ is unique by a known result \parencite[][\S2]{theoryofvariationalinequalities}.
\begin{theorem} \label{demanddecreasingimpliesunique}
If $D_c$ is strictly decreasing in $p_c$ and a market equilibrium $p^*$ exists, then the equilibrium is unique.
\end{theorem}

Combining this theorem with the argument above establishing the existence of equilibrium (\S\ref{asafixedpoint}) yields the following theorem:
\begin{theorem} \label{fullsupportimpliesunique}
If $\eta$ has full support, then the admissions market equilibrium exists and is unique.
\end{theorem}
This is Theorem 1 of Azevedo and Leshno \parencite*{supplydemandfw}, where an alternative proof is given.

\subsubsection{Convex optimization problem}
Suppose that the excess supply function $F$ defines a conservative vector field. This means that there exists a potential function (Lyupanov function) $\Phi(p)$ whose gradient is $F$:
\[\exists \Phi: \nabla_p \Phi = F\]
Such a potential function does not necessarily exist for every excess supply function. In fact, in the market considered in the second portion of this article (\S\ref{singlescoremodel}), the Jacobian of $F$ is asymmetric, which implies that $F$ is \emph{not} a conservative vector field. 

However, supposing $\Phi$ exists, the equilibrium can be found by solving the following concave maximization problem:
\begin{align*}
\text{minimize} \quad \Phi(p) \qquad \text{subject to} \quad  p \geq 0 
\end{align*}
As the feasible set has nonempty interior, Slater's condition holds, and the optimal solution $(p^*, \lambda^*)$ satisfies the following KKT conditions:
\begin{align*}
&F_c - \lambda^* = 0 & \text{(stationarity)}\\
&p^* \geq 0, ~~ \lambda^* \geq 0  & \text{(primal, dual feasibility)}\\
&\lambda^{*T} p^*=0  & \text{(complementarity)}
\end{align*}
Eliminating the dual variables $\lambda^*$ yields the nonlinear complementarity problem above (equation \eqref{nonlinearcompprob}); hence, $p^*$ is an equilibrium. Moreover, observe that the t\^{a}tonnement procedure of Algorithm \ref{admissionseqtatalgo} is a projected gradient ascent algorithm for this convex program, and vice-versa. 

\subsection{Optimization tasks}
With the equivalence results established above in hand, we can expand our understanding of admissions markets to encompass a range of optimization tasks that span a variety of realistic scenarios. One example is the canonical school-choice problem. Given $\eta$ and the capacity vector $q$, we must compute a stable matching $\mu$. It suffices to find the equivalent cutoff vector $p$, as discussed above (\S\ref{admissionseqtatalgo}). Another optimization task is the \emph{inverse optimization} problem. Given the cutoffs $p$ and demand $D$, we try to infer information about $\eta$ such as the overall preferability of each school or the joint distribution of students' scores. This task requires many simplifying assumptions, because the number of student distributions that could induce a given stable assignment is typically infinite. In the second half of this article, I turn to an example of a nonatomic admissions market in which both of these problems can be solved efficiently.

\pagebreak

\section{Single-score model with multinomial logit student preferences} \label{singlescoremodel}
In this study, I consider a special kind of admissions market that has not received much attention in the school-choice literature but approximates the admissions procedure used in many systems around the world. In this market, all schools have the same preference order, and students' preference orders are determined by the multinomial logit (MNL) choice model.

The primary reason for choosing this market is that it admits an expression for the demand that is invertible in the other parameters, allowing us to compute the equilibrium cutoffs analytically. We can also efficiently compute the gradient of the market parameters with respect to one another both in and out of equilibrium, which enables an interesting comparative analysis of the incentives available to schools under unconstrained school choice and when the market is confined to equilibrium by a deferred acceptance mechanism. Also, when the demand and cutoff vectors are known, one can solve for the preferability parameters, which yields a novel method of ranking schools' popularity and modeling their demand curves.

A single-score system may arise in one of several real-world scenarios. The most obvious case is when government regulations require schools to admit students solely on the basis of a standardized test. Alternatively, when students are scored using various dimensions of student characteristics such as test scores, GPA, and the quality of their letters of recommendation, it is common for these various dimensions to tightly correlate. If so, then principle component analysis can be used to determine a composite score whose order approximates the ordering of students at each university. Finally, in many public school systems, schools have \emph{no} preference order over the students; instead students take turns picking their favorite school in an order determined by random lottery, or (equivalently) the single tiebreaking mechanism is used to generate schools’ preference lists and the assignment of students to schools is computed using student-proposing DA \parencite[][]{whatmatters}. In this situation, the random numbers induce a single distribution of scores.

The MNL choice model represents a compromise between realism and computational tractability. In the general nonatomic school-choice problem, there are $|C|!$ possible preference orderings, and student preferences must be encoded as a probability vector of this length. A simple way to reduce this complexity is to choose a few ``representative'' preference lists, but this fails to account for the exponential number of ways in which an individual student may exchange the place of two schools within a primary list. In contrast, the MNL choice model assigns nonzero preferability to all possible preference lists while requiring only a single parameter for each school, and its parameters can be fitted via a number of known survey methodologies.\footnote{An interesting direction of future research would be to attempt to fit MNL parameters to student preference lists in a jurisdiction that uses deferred acceptance, like New York City.} The MNL choice model can also emulate to arbitrary precision the situation in which every student's preference list is \emph{identical}, by letting each school's preferability parameter differ from the next by a large order of magnitude.

\subsection{Model description} 

In this section I describe the single-score model with multinomial logit preferences, derive a closed-form expression for the demand function, and show that the demand is piecewise linear continuous and each school's demand $D_c$ is strictly decreasing in $p_c$. 

\subsubsection{Characterization of $\eta$}
To characterize $\eta$, we must describe both how schools rank students, and how students rank schools. In this model, all schools share the same ranking over the students. Since there are no ties, assume without loss of generality that the scores are uniformly distributed on the interval $[0,1]$. 

As for students' choice of school, this model assumes students use MNL choice to derive their preference lists. Each school has a preferability parameter $\delta_c \in \mathbb{R}$. Letting $C^\# \subseteq C$ denote set of schools to which a given student is admitted, she chooses to attend school $c \in C^\#$ with probability
\[\frac{\exp \delta_c}{\sum_{d \in C^\#} \exp \delta_d}\]
For convenience, let $\gamma_c \equiv \exp \delta_c > 0$ and $\Gamma = \sum_c \gamma_c$. Since the equation is homogeneous in $\gamma$, we may assume without loss of generality that $\Gamma = 1$; however, I will resist this assumption, since in a large market, taking a larger $\Gamma$-value can yield more legible parameters. 

Observe that in the single-score model with MNL choice, $\eta$ does \emph{not} have full support, because the probability of having different scores at any two schools is zero. Nonetheless, the algebraic analysis below reveals that the equilibrium is unique.

\subsubsection{Demand function}
Let us determine the demand function $D(\gamma, p)$ for the single-score model with MNL student preferences. First, sort the schools by cutoff, so that
\[p_1 \leq p_2 \leq \dots \leq p_{|C|}\]
Ties may be broken arbitrarily, as discussed below. Since getting into school $c$ implies getting into any school whose cutoff is less than or equal to $p_c$, there are only $|C| + 1$ possible consideration sets for each student, as follow.
\begin{center}
\begin{tabular}{lll}
\textbf{Symbol} & \textbf{Consideration set} & \textbf{Probability} \\ \hline
$C_{[0]}$    & $\varnothing$    & $p_1$                  \\
$C_{[1]}$    & $\left\{ c_1 \right\}$    & $p_2 - p_1$               \\
$C_{[2]}$    & $\left\{ c_1, c_2 \right\}$    & $p_3 - p_2$               \\
$\vdots$ & $\vdots$ & $\vdots$ \\
$C_{[|C| - 1]}$           & $\left\{ c_1, \dots, c_{|C| - 1} \right\}$     & $p_{|C|} - p_{|C|-1}$             \\
$C_{[|C|]}$           & $\left\{ c_1, \dots, c_{|C|} \right\}$     & $1 - p_{|C|}$                 
\end{tabular}
\end{center}
Hence, the demand for school $c$ is the sum of the number of students with each of these consideration sets who choose to attend $c$. Letting $p_{|C|+1} \equiv 1$, the demand function is as follows.
\begin{equation}D_c = \mathlarger{\mathlarger{\sum}}_{d=c}^{|C|} 
\underbrace{\frac{\exp{\delta_c}}{ \sum_{i=1}^d \exp{\delta_i}}}_{\substack{\text{probability} \\ \text{of choosing } c \\ \text{from }C_{[d]}}} 
\overbrace{\left(p_{d+1} - p_{d}\right)}^{\substack{\text{probability of}\\ \text{having consideration} \\ \text{set }C_{[d]}}} 
\label{mnlonetestdemand}\end{equation}

\subsubsection{Continuity and piecewise linearity of the demand function} \label{continuityandpiecewiselinearity}
$D$ is \emph{continuous} in $p$. To see this, expand the equation above:
\begin{gather}
\begin{aligned}D_c &= \gamma_c \left[
\left( \frac{-1}{\sum_{i=1}^c \gamma_i}\right) p_c
+ \left(\frac{1}{\sum_{i=1}^{c} \gamma_i} - \frac{1}{\sum_{i=1}^{c+1} \gamma_i} \right) p_{c+1}
\right. \\ &\left.
\quad + \cdots
+ \left(\frac{1}{\sum_{i=1}^{|C|-1} \gamma_i} - \frac{1}{\sum_{i=1}^{|C|} \gamma_i}\right) p_{|C|}
+ \frac{1}{\sum_{i=1}^{|C|} \gamma_i}
\right]
\end{aligned}
\end{gather}
Since $D$ is linear in any neighborhood where the order of cutoffs is unambiguous, the only opportunity for discontinuity occurs when two or more cutoffs are equal. Thus, it suffices to show that the value of $D_c$ is independent of how ties among the $p_c$ are broken. Suppose that $p_j = \dots = p_{j+n} = \tilde p$ for some $j > c$. Then (dividing by $\gamma_c$ for legibility)
\begin{gather}
\begin{aligned}
\frac{D_c}{\gamma_c} &= \cdots
+ \left(\frac{1}{\sum_{i=1}^{j-1} \gamma_i} - \frac{1}{\sum_{i=1}^{j} \gamma_i} \right) p_{j}
+ \left(\frac{1}{\sum_{i=1}^{j} \gamma_i} - \frac{1}{\sum_{i=1}^{j+1} \gamma_i} \right) p_{j+1}
 \\ &\quad + \cdots
+ \left(\frac{1}{\sum_{i=1}^{j+n} \gamma_i} - \frac{1}{\sum_{i=1}^{j+n+1} \gamma_i}\right) p_{j+n}
+ \cdots \\
&= \cdots
+ \left(\frac{1}{\sum_{i=1}^{j-1} \gamma_i} - \cancel{\frac{1}{\sum_{i=1}^{j} \gamma_i}} \right) \tilde p
+ \left(\cancel{\frac{1}{\sum_{i=1}^{j} \gamma_i}} - \cancel{\frac{1}{\sum_{i=1}^{j+1} \gamma_i}} \right) \tilde p
\\ &\quad  + \cdots
+ \left(\cancel{\frac{1}{\sum_{i=1}^{j+n} \gamma_i}} - \frac{1}{\sum_{i=1}^{j+n+1} \gamma_i}\right) \tilde p
+ \cdots \\
&= \cdots
+ \left(\frac{1}{\sum_{i=1}^{j-1} \gamma_i} - \frac{1}{\sum_{i=1}^{j+n+1} \gamma_i}\right) \tilde p
+ \cdots
\end{aligned}
\end{gather}
The internal sums that depend on the order of the indices $j \dots j+n$ cancel out; hence, they may be arbitrarily reordered without changing the value of $D_c$. Similar canceling shows that the demand does not vary under tiebreaking when $c$ itself is involved in a tie. Hence, $D$ is continuous in $p$. 

The expansion above also allows us to see that the demand vector is defined by the matrix equation
\begin{equation}D = A p + \frac{1}{\Gamma}\gamma \label{demandmatrixeq}\end{equation}
where $A\in \mathbb{R}^{|C| \times |C|}$ is the triangular matrix with
\begin{align} \label{Adef}
A_{ij} &\equiv \begin{cases}
0, & i > j \\
-\gamma_i \left(\frac{1}{ \sum_{k=1}^i \gamma_k}\right), & i=j \\
\gamma_i \left( \frac{1}{\sum_{k=1}^{j-1} \gamma_k} -  \frac{1}{\sum_{k=1}^{j} \gamma_k}\right), & i<j \\
\end{cases} \\[.8em]
\iff A &= \begin{bmatrix}
\gamma_1 \left( \frac{-1}{\gamma_1} \right) & \gamma_1 \left(\frac{1}{\gamma_1} - \frac{1}{\gamma_1 + \gamma_2} \right) & \gamma_1 \left(\frac{1}{\gamma_1 + \gamma_2} - \frac{1}{\gamma_1 + \gamma_2 + \gamma_3} \right) & \cdots &  \gamma_1 \left(\frac{1}{\sum_{i=1}^{|C| - 1}\gamma_i} - \frac{1}{\Gamma}  \right)  \\
 & \gamma_2 \left( \frac{-1}{\gamma_1 + \gamma_2} \right) & \gamma_2 \left(\frac{1}{\gamma_1 + \gamma_2} - \frac{1}{\gamma_1 + \gamma_2 + \gamma_3} \right) & \cdots &  \gamma_2 \left(\frac{1}{\sum_{i=1}^{|C| - 1}\gamma_i} - \frac{1}{\Gamma} \right)  \\
 &  & \gamma_3 \left( \frac{-1}{\gamma_1 + \gamma_2 + \gamma_3} \right) & \cdots &  \gamma_3 \left(\frac{1}{\sum_{i=1}^{|C| - 1}\gamma_i} - \frac{1}{\Gamma} \right)  \\
 & & & \ddots & \vdots \\
 &  &  &  &  \gamma_{|C|} \left(\frac{1}{\sum_{i=1}^{|C| - 1}\gamma_i} -\frac{1}{\Gamma}  \right)  \\
\end{bmatrix}\end{align}

Since $\gamma > 0$, $A$ is invertible. This $A$ will reappear throughout the analysis. 

The matrix $A$ depends on the order of the $p_c$ values, so the demand function is \emph{piecewise linear} in $p$.\footnote{In the context of an iterative schema such as the t\^{a}tonnement process simulated in Figure \ref{tat-iter-cutoff} below, instead of sorting $p$ itself, it is often simpler to permute the rows and columns of $A$ according to the inverse of the permutation that sorts $p$.} Because the main diagonal of $A$ is strictly negative, the demand at each school $c$ is strictly decreasing in $p_c$. By Theorem \ref{demanddecreasingimpliesunique}, it follows that that the equilibrium is unique.

\subsubsection{Appeal function}
An interesting indicator from Azevedo and Leshno \parencite*{supplydemandfw} is the \emph{appeal} of a school's entering class, or the integral of scores over the set of admitted students. The average score of a student with consideration set $C_{[d]}$ is $\frac{1}{2}\left(p_{d+1} + p_d\right)$, so the appeal at $c$ is
\begin{align}
L_c &= \sum_{d=c}^{|C|} 
\underbrace{\frac{\exp{\delta_c}}{ \sum_{i=1}^d \exp{\delta_i}}}_{\substack{\text{probability} \\ \text{of choosing } c \\ \text{from }C_{[d]}}} 
\overbrace{\left(p_{d+1} - p_{d}\right)}^{\substack{\text{probability of}\\ \text{having consideration} \\ \text{set }C_{[d]}}} 
\underbrace{\frac{1}{2}\left(p_{d+1} + p_{d}\right)}_{\substack{\text{avg. score of students}\\ \text{with consideration} \\ \text{set }C_{[d]}}}
=\frac{1}{2}\sum_{d=c}^{|C|} 
\frac{{\gamma_c}}{ \sum_{i=1}^d {\gamma_i}} 
\left(p_{d+1}^2 -  p_{d}^2\right)
\end{align}
By comparison with the expression for $D$, the appeal vector is given by 
\[L = \frac{1}{2} A p.^2 + \frac{1}{2\Gamma} \gamma\]
where the notation $p.^2 = (p_1^2, \dots, p_{|C|}^2)$ represents the entrywise square of $p$.

The appeal of the entering class is not necessarily the school's objective function, because schools may value an abstract notion of selectivity or students' tuition dollars higher than this value. Moreover, interpreting $L_c$ as utility imputes cardinal information to the scores by assuming that each school regards two students scoring $0.2$ as equivalent to one student scoring $0.4$. A more accurate model of school utility, not considered here, should associate with each school a valuation function of scores in the unit interval, and compute the integral of this valuation function over the distribution of assigned students. Such a model would induce an different competitive equilibrium.

\subsection{Computing the equilibrium}
In the market under consideration, the equilibrium conditions are as follows:
\begin{gather} \label{ssmnleqconds}
\begin{aligned}
D = A p + \frac{1}{\Gamma}\gamma &\leq q \\
D_c = A_{c.} p + \frac{1}{\Gamma} \gamma_c &= q_c, \quad \forall c: p_c > 0
\end{aligned}
\end{gather}
As I will now show, it turns out that at equilibrium, the order of the school cutoffs is determined by the order of the \emph{competitiveness ratios} $\gamma_c / q_c$. This fact enables us to compute the equilibrium directly by solving a linear system. Below, the positive part operator $x^+$ works elementwise on its argument $x$. That is, $(x^+)_i \equiv \max\{0, x_i\}$.

\begin{theorem} \label{cutoffsortationthm}
Without loss of generality, suppose that $\frac{\gamma_1}{q_1} \leq \dots \leq \frac{\gamma_{|C|}}{q_{|C|}}$. Then $\hat p_1 \leq \cdots \leq \hat p_{|C|}$, and
\[\hat p \equiv \left[A^{-1} (q - \frac{1}{\Gamma} \gamma) \right]^+\]
is the market equilibrium in the single-score, MNL choice model. Moreover, the equilibrium is unique. 
\end{theorem} 

\begin{proof}
I show the following statements:
\begin{enumerate}
\item $\hat p$ satisfies $\hat p_1 \leq \cdots \leq \hat p_{|C|}$. This means that the demand at $\hat p$ is given by the expression $A \hat p + \frac{1}{\Gamma}\gamma$ (which only holds if $\hat p$ is sorted).
\item $\hat p$ satisfies the equilibrium conditions given in equation \eqref{ssmnleqconds}.
\end{enumerate}
For convenience, let $\bar p \equiv A^{-1} (q - \frac{1}{\Gamma} \gamma) $, so that $\hat p = \bar p^+$. 
\begin{enumerate}
\item Pick any school $c < |C|$. It suffices to show that $\bar p_{c+1} - \bar p_{c} \geq 0$. The inverse of $A$ is
\begin{equation} \label{Ainv}
A^{-1} = \begin{bmatrix}
\frac{-1}{\gamma_1}\left( \gamma_1 \right) & -1 & -1 &\cdots & -1 \\
 & \frac{-1}{\gamma_2}\left( \gamma_1 + \gamma_2 \right) & -1 &\cdots & -1 \\
 & & \frac{-1}{\gamma_2}\left( \gamma_1 + \gamma_2 + \gamma_3 \right) &\cdots & -1 \\
 &  &  & \ddots & \vdots \\
 & & & &  \frac{-1}{\gamma_{|C|}} \Gamma \\
\end{bmatrix}
\end{equation}
It is not difficult to verify that
\begin{align} \label{barpissorted}
\bar p_{c+1} - \bar p_{c}
&= \left[A^{-1} (q - \frac{1}{\Gamma} \gamma) \right]_{c+1} - \left[A^{-1} (q - \frac{1}{\Gamma} \gamma) \right]_{c} = \left(\sum_{j=1}^c \gamma_j \right) \left(\frac{q_c}{\gamma_c} - \frac{q_{c+1}}{\gamma_{c+1}}\right) \geq 0
\end{align}
which follows from the assumption that $\gamma_c / q_c \leq \gamma_{c+1} / q_{c+1}$. Hence, $\bar p$ is sorted, and so is $\hat p$. 


\item The demand at $\hat p$ is $D = A \hat p + \frac{1}{\Gamma}\gamma$. Hence
\begin{align*}
\hat p = A^{-1} (D - \frac{1}{\Gamma} \gamma) = \bar p^+ &\geq \bar p = A^{-1} (q - \frac{1}{\Gamma} \gamma) \\
\implies \quad A^{-1} D &\geq A^{-1} q \\
\implies \quad D &\leq q
\end{align*}
The final statement follows from the fact that $A^{-1}$ is triangular and its nonzero entries are strictly negative. This establishes the capacity condition. 

Now, I need to show that the demand equals the capacity when $\hat p_c > 0$. Let $b$ denote the first school with a nonzero cutoff. That is, $\hat p_1 = \dots = \hat p_{b-1} = 0$, and $0 < \hat p_b \leq p_{b+1} \leq \dots \leq \hat p_{|C|}$. Then the demand at $\hat p$ may be written
\begin{gather}\begin{aligned} \label{demandatphat}
D &= A \hat p + \frac{1}{\Gamma}\gamma \\
&= \sum_{i=1}^{|C|} A_{.i} \hat p_i + \frac{1}{\Gamma}\gamma  \\
&= \sum_{i=1}^{|C|} A_{.i} \left[A^{-1} \left(q - \frac{1}{\Gamma}\gamma\right) \right]_i^+ + \frac{1}{\Gamma}\gamma  \\
&= \sum_{j=b}^{|C|} A_{.j} \left[A^{-1} \left(q - \frac{1}{\Gamma}\gamma\right) \right]_j + \frac{1}{\Gamma}\gamma  \\
&= \left[\sum_{j=b}^{|C|} A_{.j} A_{j.}^{-1} \right] \left(q - \frac{1}{\Gamma}\gamma\right) + \frac{1}{\Gamma}\gamma  \\
&= \begin{bmatrix}
0_{b \times b} & T_{b \times (|C| - b)} \\
0_{(|C| - b) \times b} & I_{|C| - b} \\
\end{bmatrix} \left(q - \frac{1}{\Gamma}\gamma\right) + \frac{1}{\Gamma}\gamma  \\
\end{aligned}\end{gather}
where
\begin{equation} \label{Tdef}
T = \begin{bmatrix}
\frac{-\gamma_1}{\sum_{i=1}^{b-1} \gamma_i} & \cdots & \frac{-\gamma_1}{\sum_{i=1}^{b-1} \gamma_i} \\
\vdots & \cdots & \vdots \\
\frac{-\gamma_{b-1}}{\sum_{i=1}^{b-1} \gamma_i} & \cdots & \frac{-\gamma_{b-1}}{\sum_{i=1}^{b-1} \gamma_i}
\end{bmatrix}\end{equation}
For the schools with $\hat p_c > 0$, the demand is
\begin{align} \label{demand-pc-gt-zero}
D_c &=
\begin{bmatrix}
0& I
\end{bmatrix}_{c.} \left(q - \frac{1}{\Gamma}\gamma\right) + \frac{1}{\Gamma}\gamma
= q_c
\end{align}
\end{enumerate}
Hence, the stability criterion holds, and $\hat p$ is an equilibrium.
\end{proof}

For reference, for the schools with $\hat p_c = 0$, the demand at equilibrium is 
\begin{align} \label{demand-pc-eq-zero}
D_c &=
\begin{bmatrix}
0& T
\end{bmatrix}_{c.} \left(q - \frac{1}{\Gamma}\gamma\right) + \frac{1}{\Gamma}\gamma  
= \frac{-\gamma_c}{\sum_{i=1}^{b-1} \gamma_i} \sum_{j=b}^{|C|} \left(q_j - \frac{1}{\Gamma}\gamma_j\right)  + \frac{1}{\Gamma}\gamma_c \leq q_c
\end{align}

With these results in hand, I turn to a comparative analysis of the incentives that two different assignment mechanisms provide to schools in this market.

\subsection{Incentive gradients under decentralized assignment mechanisms}
In this section, I consider the incentives available to schools in an unconstrained market in which schools have no capacity constraints. Then, in the following section, I consider these incentives in a centralized market that always produces a stable matching. In principle, schools can have any objective function, but the analysis below assumes that each school's utility is increasing in its cutoff, its demand, and its appeal. Under decentralized assignment, both the cutoff and quality can be interpreted as variables within each school's control; under centralized assignment, schools can only affect their quality. Hence, the set of derivatives that afford a meaningful interpretation differs somewhat between the centralized and decentralized cases. 

In the decentralized market, schools are obligated to allow students to enroll if offered admission; equivalently, there is ample room for any student at any school, as long as she meets its admissions standards. Thus, each school can set its own cutoff $p_c$ in reflection of its own admissions goals. And, to the extent it can, each school can try to increase its preferability $\gamma_c$ by advertising, updating its curriculum, and so on. The present discussion considers the effect of these moves on the inputs to the school's objective function. 

\subsubsection{Cutoff effects} \label{unconstrainedcutoffeffects}
The response of the demand to a change in cutoffs is the Jacobian of the demand function:
\[\mathbf{J}_p D = A \]
The diagonal is negative, meaning that each school's demand is decreasing in its cutoff, as expected. The entries above the diagonal are positive, while those below the diagonal are zero. This means that each school $c$'s demand is increasing in the cutoffs of the \emph{more-selective} schools, but the cutoffs of \emph{less-selective schools} have no local effect on the demand at $c$.

Intuitively, this means that if all schools are equally preferable, a highly selective school has more market power than the others: If it increases its cutoff, it will cause many students to move onto another school. On the other hand, a school $c'$ that is less preferable than $c$ cannot affect $D_c$'s demand by changing its own cutoff, because any student currently admitted to $c$ was already admitted to $c'$, and chose $c$ instead. 

Observe also that $-1 = A_{11} < A_{22} < \dots < A_{|C||C|} < 0$. This says that the school with the most generous cutoff has the most power to increase its demand with a marginal decrease in $p_c$. Intuitively, this is because a student who gets into a school with a large cutoff gets into \emph{many} schools, so competition for this student is fiercer than for a student whose options are already limited by a low score. 

Next, consider how the entering classes' appeal responds to a change in cutoffs:
\[\mathbf{J}_p L = A\operatorname{diag}(p)\]
For $p_c > 0$, the cutoff effect on appeal has the same direction as the cutoff effect on demand. Intuitively, this suggests that if a school's goal is to maximize the appeal of its entering class, it will tend to try to lower its score cutoffs as much as it can, subject to constraints on its total demand. However, the magnitude of the incentive increases when $p_c$ is higher. This tends to counteract the market power effect described above: A school with a low cutoff has the power to attract more marginal students, but does so with little overall effect on the aggregate appeal of its entering class. In the extreme case, when $p_c = 0$, the appeal associated with a marginal student is exactly zero. The competitive equilibrium of the admissions market when each school's appeal function is its utility function (which is distinct from the notion of equilibrium considered here) thus always has $p^* = 0$. 

The derivatives given above are well-defined when the cutoffs are totally ordered. However, an edge case occurs when there is a tie among the cutoffs; then the subdifferential set is given by the convex hull of the Jacobians associated with the possible permutations of $p$. 

\subsubsection{Quality effects} \label{unconstrainedqualityeffects}
Differentiate the demand with respect to $\gamma$ to obtain the effect of a marginal change in quality.
\begin{equation} \label{jac-gamma-demand-uncons}
\left(\mathbf{J}_\gamma D \right)_{c\hat c} =
\frac{\partial}{\partial\gamma_{\hat c}} D_c = \begin{cases}
\sum_{d=c}^{|C|} \frac{-\gamma_c}{\left(\sum_{i=1}^{d} \gamma_i \right)^2} \left(p_{d+1} - p_d \right), & \hat c < c \\
\sum_{d=c}^{|C|} \frac{1}{\sum_{i=1}^{d} \gamma_i}
    \left( 1 - \frac{\gamma_c}{\sum_{i=1}^{d}\gamma_i }\right)
    \left(p_{d+1} - p_d \right), & \hat c = c\\
\sum_{d=\hat c}^{|C|} \frac{-\gamma_c}{\left(\sum_{i=1}^{d}\gamma_i \right)^2} \left(p_{d+1} - p_d \right), & \hat c > c
\end{cases}
\end{equation}
(Note that the $\hat c > c$ and $ \hat c < c$ cases differ in the outer sum's starting index.) The demand for $c$ is predictably decreasing in the quality of the other schools and increasing in $\gamma_c$. This Jacobian and a partial graph of schools’ demand curves in a fictional market are given in Figure \ref{vary-gamma-demand}. 

A similar picture emerges when we differentiate the appeal with respect to $\gamma$:
\begin{equation}
\left(\mathbf{J}_\gamma L \right)_{c\hat c} =
\frac{\partial}{\partial\gamma_{\hat c}} L_c = \begin{cases}
\frac{1}{2}\sum_{d=c}^{|C|} \frac{-\gamma_c}{\left(\sum_{i=1}^{d}\gamma_i \right)^2} \left(p_{d+1}^2 - p_d^2 \right), & \hat c < c \\
\frac{1}{2}\sum_{d=c}^{|C|} \frac{1}{\sum_{i=1}^{d} \gamma_i}
    \left( 1 - \frac{\gamma_c}{\sum_{i=1}^{d} \gamma_i}\right)
    \left(p_{d+1}^2 - p_d^2 \right), & \hat c = c\\
\frac{1}{2}\sum_{d=\hat c}^{|C|} \frac{-\gamma_c}{\left(\sum_{i=1}^{d} \gamma_i \right)^2} \left(p_{d+1}^2 - p_d^2 \right), & \hat c > c
\end{cases}\end{equation}
By the same procedure used to show the continuity of $D$ above, it is possible to show that the quality effects are continuous across tiebreaking permutations of $p$.

\subsection{Comparative statics at equilibrium} \label{compstateq}
Now, I analyze the incentives available to schools when the market is constrained to equilibrium, for example, by a centralized admissions process that uses a DA algorithm to produce a stable matching, or by the assumption that the cutoffs in a decentralized market will quickly correct toward equilibrium after a few admissions cycles. Throughout this section, I assume that schools are indexed in ascending order by the competitiveness ratios $\gamma_c / q_c$. The quantities derived here were proposed by Azevedo and Leshno \parencite*{supplydemandfw}, but not computed analytically.

\subsubsection{Quality effects at equilibrium} \label{qualityeffectsateq}
First, I will focus on the effect of a marginal change in quality on the allocation of students at equilibrium. Since, in theory, schools have the power to change their own quality by investing in their programs or marketing, the analysis below enables us to quantify the extent to which these investments are ``worth it'' with respect to the school's interest in maintaining high admissions standards or increasing its demand. 

First, I provide yet another expression for the equilibrium cutoff vector $\hat p$, which can be verified by expanding the equation given in Theorem \ref{cutoffsortationthm}. $\hat p_c = \bar p_c^+$, where
\begin{equation} \label{yetanothereqcutoff}
\bar p_c = 
\frac{1}{\gamma_c} \left(\frac{\gamma_c}{\Gamma} - q_c\right) \sum_{i=1}^{c} \gamma_i 
+ \sum_{j=c+1}^{|C|} \left( \frac{\gamma_j}{\Gamma} - q_j \right)
\end{equation}
and, in the $c = |C|$ case, I take $\sum_{j=|C|+1}^{|C|} \left( \frac{\gamma_j}{\Gamma} - q_j \right)= 0$. This assumes the schools are indexed in ascending order by the competitiveness ratios $\gamma_c / q_c$. 

Differentiating the optimal cutoffs with respect to the quality and simplifying, we have
\begin{equation}\label{jac-gamma-p}
\left(\mathbf{J}_\gamma \hat p\right)_{c\hat c} =
\frac{\partial}{\partial\gamma_{\hat c}} \hat p_c = \begin{cases}
0, & \bar p_c < 0 \\
\text{undefined}, & \bar p_c = 0 \\
 - \frac{q_c}{\gamma_c}, & \bar p_c > 0 \text{ and }\hat c < c \\
\frac{q_c}{\gamma_c^2} \sum_{i=1}^{c-1} \gamma_i, & \bar p_c > 0 \text{ and }\hat c = c\\
0, & \bar p_c > 0 \text{ and }\hat c > c
\end{cases}
\end{equation}
In the $c=1$ case, again interpret the empty set as summing to zero: $\sum_{i=1}^{0} \gamma_i = 0$. This means that the entry in the top left is always zero. The Jacobian is lower triangular: any change in the quality of a school whose competitiveness ratio is already higher than that of $c$ induces no change in the cutoff at $c$. This calculation is applied to a fictional market and verified graphically in Figure \ref{vary-gamma-cutoff}.

Applying the chain rule to the demand at equilibrium $D = A \hat p + \frac{1}{\Gamma} \gamma$, and letting $b$ denote the index of the first school with a nonzero cutoff (as above), the derivative of the equilibrium demand at $c$ with respect to the quality of $\hat c$ is
\begin{equation} \label{jac-gamma-demand}
\left(\mathbf{J}_\gamma D\left(\hat p\right)\right)_{c\hat c} =
\frac{\partial}{\partial\gamma_{\hat c}} D(\hat p_c) = \begin{cases}
-\gamma_c \frac{1 - \sum_{j=b}^{|C|} q_j}{\left(\sum_{i=1}^{b-1} \gamma_i\right)^2}, & \bar p_c < 0 \text{ and }\hat c \neq c \\
\left(- \gamma_c + \sum_{k=1}^{b-1} \gamma_k\right)\frac{1 - \sum_{j=b}^{|C|} q_j}{\left(\sum_{i=1}^{b-1} \gamma_i\right)^2}, & \bar p_c < 0 \text{ and }\hat c = c\\
\text{undefined}, & \bar p_c = 0 \\
0, & \bar p_c > 0
\end{cases}
\end{equation}

Disregarding the knife-edge case in which $\bar p_c = 0$, the two derivatives above suggest that schools in the single-test model fall into one of two clear categories. For the schools for which $\bar p_c < 0$, a marginal improvement in quality increases the \emph{size} of the entering class but has no effect on its \emph{minimum score} (and, in general, the effect on the average score is small). On the other hand, for the schools for which $\bar p_c > 0$, their capacity is always filled at equilibrium, and any investment in quality yields immediate improvement in the minimum score of the entering class. If the objective functions are a combination of cutoff and demand, this analysis suggests that competition within these two broad groups of schools is close to zero-sum. Underdemanded schools compete for the finite pool of tuition dollars remaining in the market after the best students have chosen the top schools, whereas overdemanded schools compete for the top slice of the fixed distribution of student talent.

The appeal function also be differentiated in $\gamma$, although it does not admit a legible representation for an arbitrary number of schools.

\subsubsection{Capacity and population effects}
Consulting the cutoff sortation result of Theorem \ref{cutoffsortationthm}, it is easy to see that the derivative of the equilibrium cutoffs with respect to a given school's capacity is
\begin{equation}\label{jac-q-p}
\left(\mathbf{J}_q \hat p\right)_{c\hat c} =
\frac{\partial}{\partial q_{\hat c}} \hat p_c = \begin{cases}
0, & \bar p_c < 0 \\
\text{undefined}, & \bar p_c = 0 \\
A^{-1}_{c \hat c}, & \bar p_c > 0 \text{ and }\hat c < c 
\end{cases}
\end{equation}

The derivative of the demand, by inspecting equation \eqref{demandatphat}, has
\begin{equation}\mathbf{J}_q D(\hat p) =
\begin{bmatrix}
0_{b \times b} & T_{b \times (|C| - b)} \\
0_{(|C| - b) \times b} & I_{|C| - b} \\
\end{bmatrix} 
\end{equation}
where the entries of $T$ are negative as given in equation \eqref{Tdef}.

This confirms the intuitive result that only schools that are overdemanded at equilibrium can make use of excess capacity. In addition, observe that because $\mathbf{J}_q \hat p$ is upper triangular, adding capacity to a school whose competitiveness ratio is lower than that of $c$ has no marginal effect on the equilibrium cutoff at $c$. 

%


\subsection{A numerical example}
In this section, I offer a numerical demonstration of Theorem \ref{cutoffsortationthm}'s sortation result. I validate the interpretation of the equilibrium cutoffs as a stationary point of a t\^{a}tonnement process, as a market-clearing price vector, and as the limit point of stable assignments. (The demonstration of the latter two interpretations follows Azevedo and Leshno \parencite*{supplydemandfw}.) Finally, I represent the incentive results above graphically.

\begin{figure}
\begin{center}\includegraphics[width=\linewidth, ]{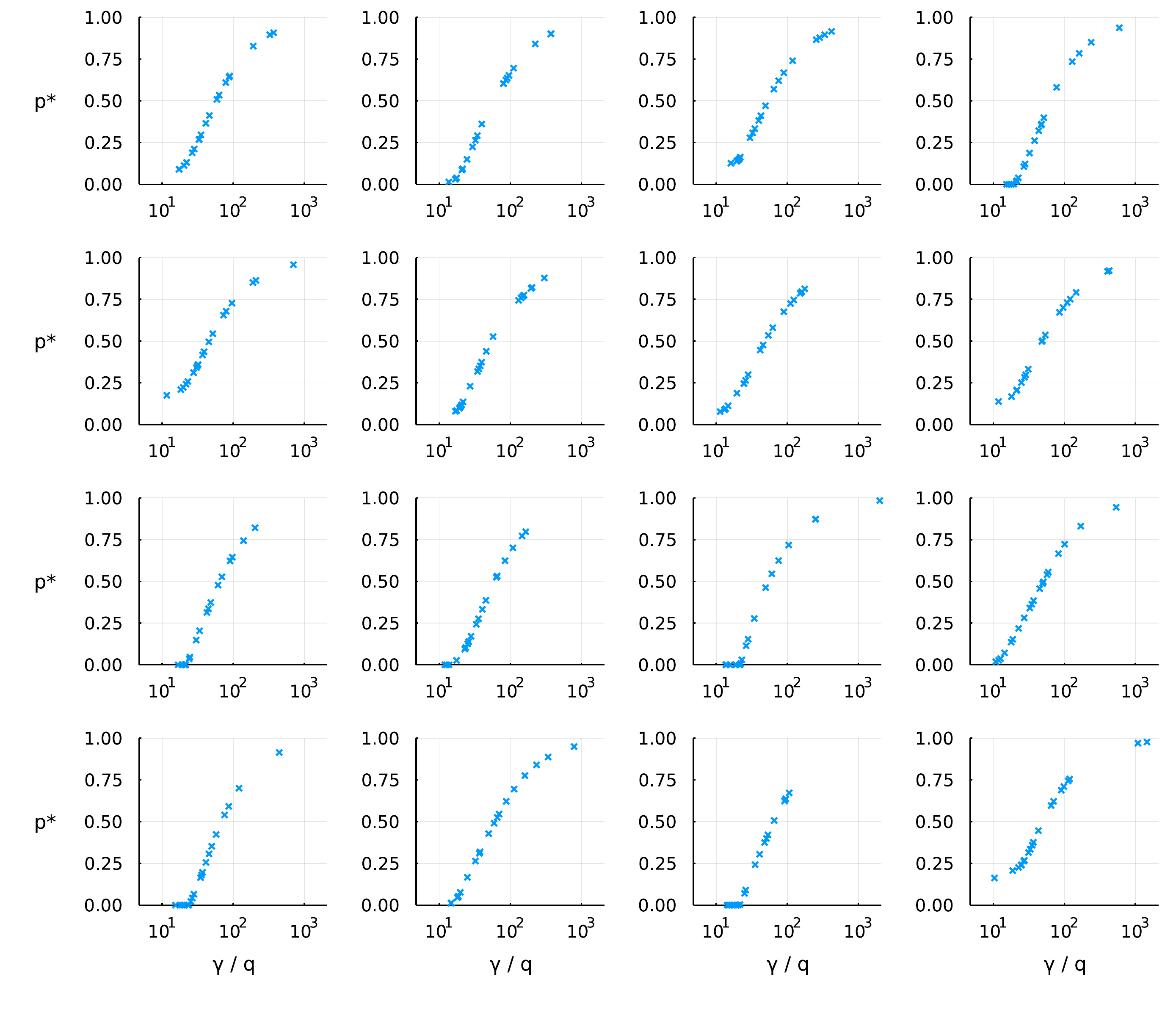}\end{center}
\captionsetup{singlelinecheck=off}
    \caption[.]{Competitiveness ratios $\gamma_c / q_c$ and equilibrium cutoffs $p_c^*$ in 16 randomly generated admissions markets, each containing 20 schools. The preferability parameters $\delta_c$ are drawn from $\operatorname{Uniform(0, 1)}$, while the capacities are drawn from $\operatorname{Uniform(0, 1/10)}$; hence, the market as a whole has a 50 percent chance of being over- or underdemanded. The figure suggests Theorem \ref{cutoffsortationthm}, which states that the order of the equilibrium cutoffs is determined by the order of competitiveness ratios.}
\label{gammaq-pstar}
\end{figure}

Figure \ref{gammaq-pstar} demonstrates the relationship between the equilibrium cutoffs $p_c^*$ and the competitiveness ratios $\gamma_c / q_c$ in sixteen randomly generated markets. Some of the markets are overdemanded, yielding $p^* > 0$, and others are underdemanded; however, the equilibrium cutoffs are always ordered according to the competitiveness ratios. As the graph indicates, the precise relationship is nonlinear and highly sensitive to variance in the market parameters. This suggests that even in the highly stylized model under consideration, it is difficult for a school to predict the effect of small perturbation in a single $\gamma_c$ or $q_c$ value on the market as a whole myopically---that is, by looking only at its entering class. Instead, and especially under a stable assignment paradigm, schools must model their demand curve in a way that accounts for the second-order effect of a change in cutoff on the consideration sets of marginal students.

\begin{figure}
\begin{center}\includegraphics[width=\linewidth, ]{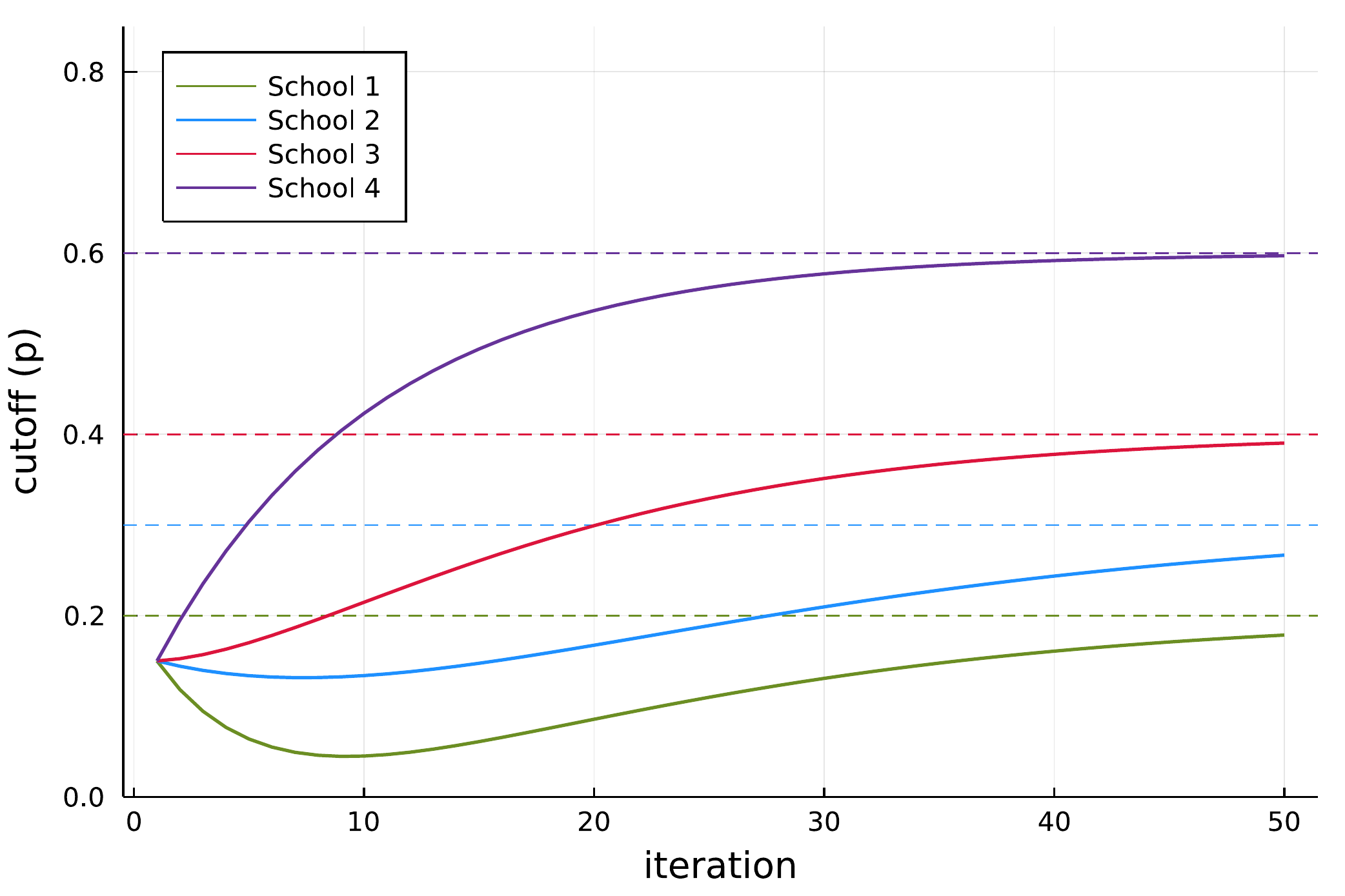}\end{center}
\captionsetup{singlelinecheck=off}
    \caption[.]{Convergence toward equilibrium in fifty iterations of the simultaneous t\^{a}tonnement process (Algorithm \ref{admissionseqtatalgo}) in the fictional nonatomic admissions market of Pallet Town, whose parameters are given in equation \eqref{pallettowndef}. The step size parameters are $\alpha = 0.2$ and $\beta = 0.01$, and the initial cutoff vector is $p^{(0)} = (0.15, 0.15, 0.15, 0.15)$. }
\label{tat-iter-cutoff}
\end{figure}

\begin{figure}
\begin{center}\includegraphics[width=\linewidth, ]{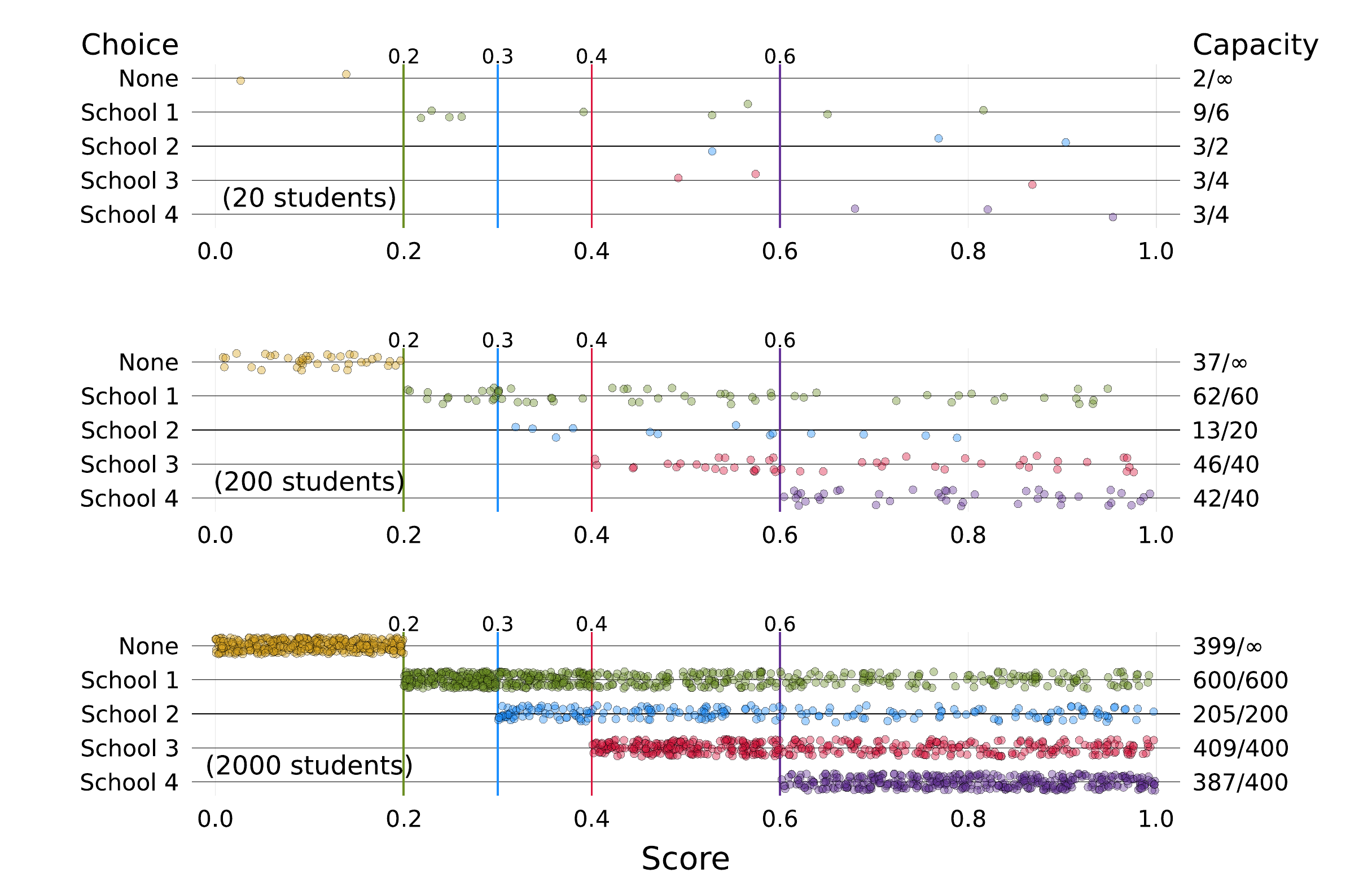}\end{center}
\captionsetup{singlelinecheck=off}
    \caption[.]{Simulation of a decentralized school-choice process in Pallet Town. A discrete sample of student preference lists and scores is drawn from $\eta$. Each school admits students whose score exceeds its equilibrium cutoff (shown as vertical lines), then each student chooses her favorite school from her consideration set. As the sample size increases, the demand at each school approximately equals its scaled capacity.}
\label{score-cutoff-choice}
\end{figure}

\begin{figure}
\begin{center}\includegraphics[width=\linewidth, ]{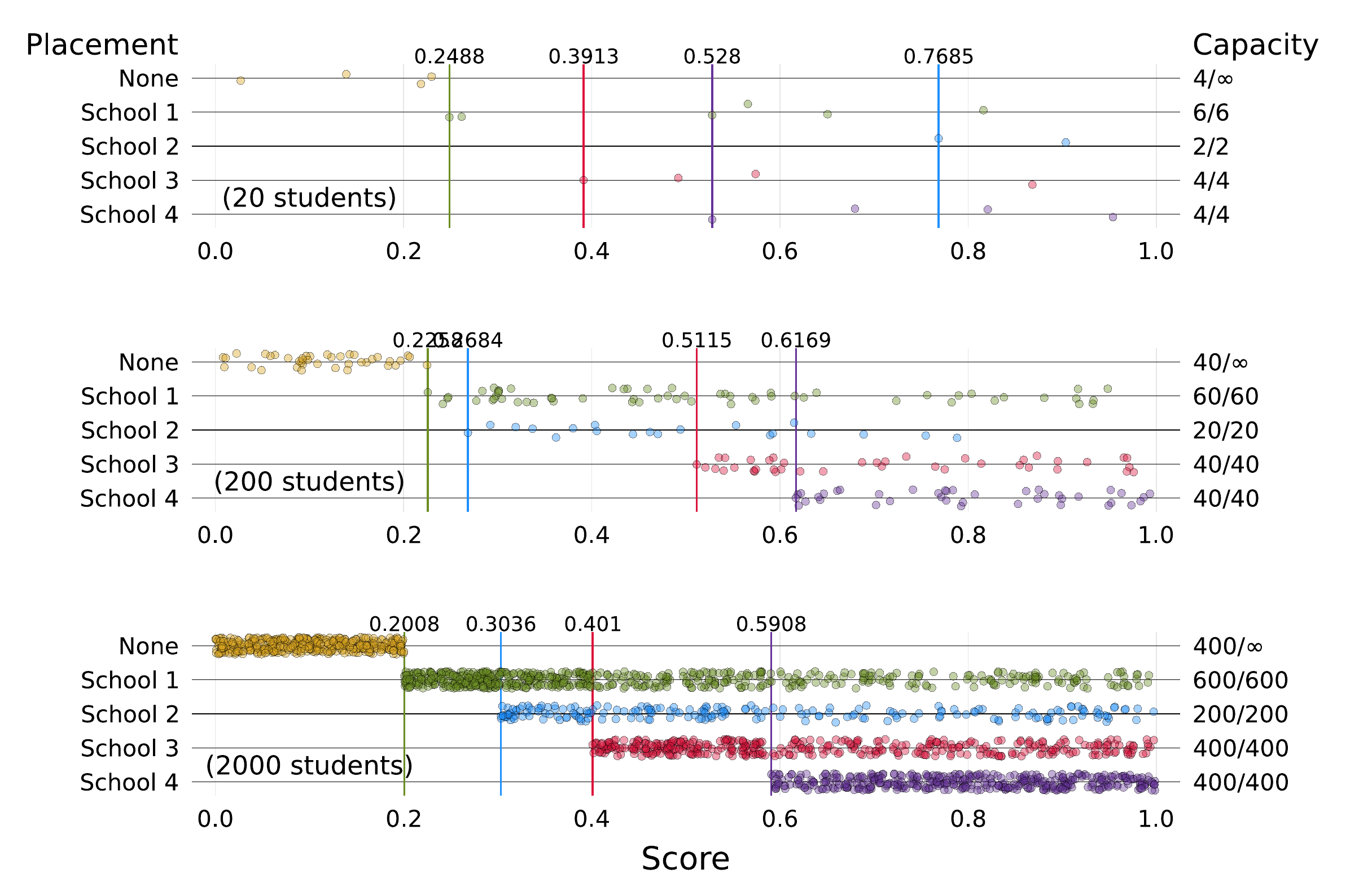}\end{center}
\captionsetup{singlelinecheck=off}
    \caption[.]{Simulation of a deferred acceptance process in Pallet Town. A discrete sample of student preference lists and scores is drawn from $\eta$. The student-proposing deferred acceptance algorithm (Algorithm \ref{studentproposingDA}) is used to compute a stable matching. The score of the least-qualified admit at each school, or the school's implied score cutoff, is computed and represented as a vertical line. Regardless of the sample size, each school fills its scaled capacity, and as the sample size increases, the implied cutoffs approach the market equilibrium. Comparison with Figures \ref{tat-iter-cutoff} and \ref{score-cutoff-choice} suggests the equivalence among stationary points of a t\^{a}tonnement process, market clearing cutoffs, and stable assignments in nonatomic admissions markets.}
\label{score-DA-placement}
\end{figure}

Figures \ref{tat-iter-cutoff} through \ref{score-DA-placement} consider a fictional admissions market called Pallet Town, which has the following parameters. I have sorted the schools by their equilibrium cutoffs.
\begin{gather} \label{pallettowndef}
\begin{aligned}
\gamma &=  \textstyle{\left(\frac{2}{12}, \frac{1}{12}, \frac{3}{12}, \frac{6}{12}\right)}\\
q &= (0.3, 0.1, 0.2, 0.2) \\
p^* &= (0.2, 0.3, 0.4, 0.6)\\
D(p^*, \gamma) &=  (0.3, 0.1, 0.2, 0.2)
\end{aligned}
\end{gather}
As the total capacity is less than one, each school fills its capacity at equilibrium.

Figure \ref{tat-iter-cutoff} shows fifty iterations of the simultaneous t\^{a}tonnement process (Algorithm \ref{admissionseqtatalgo}) applied to Pallet Town. At each iteration, the demand is computed directly by evaluating the expression derived in equation \eqref{mnlonetestdemand}. Then, the cutoff vector is adjusted in the direction of excess demand according to a predetermined, decreasing sequence of step sizes. The cutoffs converge smoothly toward $p^*$, which suggests the continuity of the demand function and the stability of the equilibrium. 

Figures \ref{score-cutoff-choice} and \ref{score-DA-placement} consider discrete approximations of the Pallet Town admissions market with samples of 20, 200, and 2000 students. In Figure \ref{score-cutoff-choice}, schools admit students according to their equilibrium cutoffs, students choose their favorite school, and schools observe their demand. When there are many students, the demand at each school approximately equals its capacity. In Figure \ref{score-DA-placement}, a stable matching of the students is computed so that each school fills its capacity exactly. When there are many students, the implied cutoffs approximately equal $p^*$. Compare Figures \ref{score-cutoff-choice} and \ref{score-DA-placement} with figure 4 of Azevedo and Leshno \parencite*{supplydemandfw}, which demonstrates the asymptotic convergence of implicit cutoffs as the number of students increases using a numerical experiment in which scores are partially correlated between schools. 

\begin{figure}
\begin{center}\includegraphics[width=\linewidth, ]{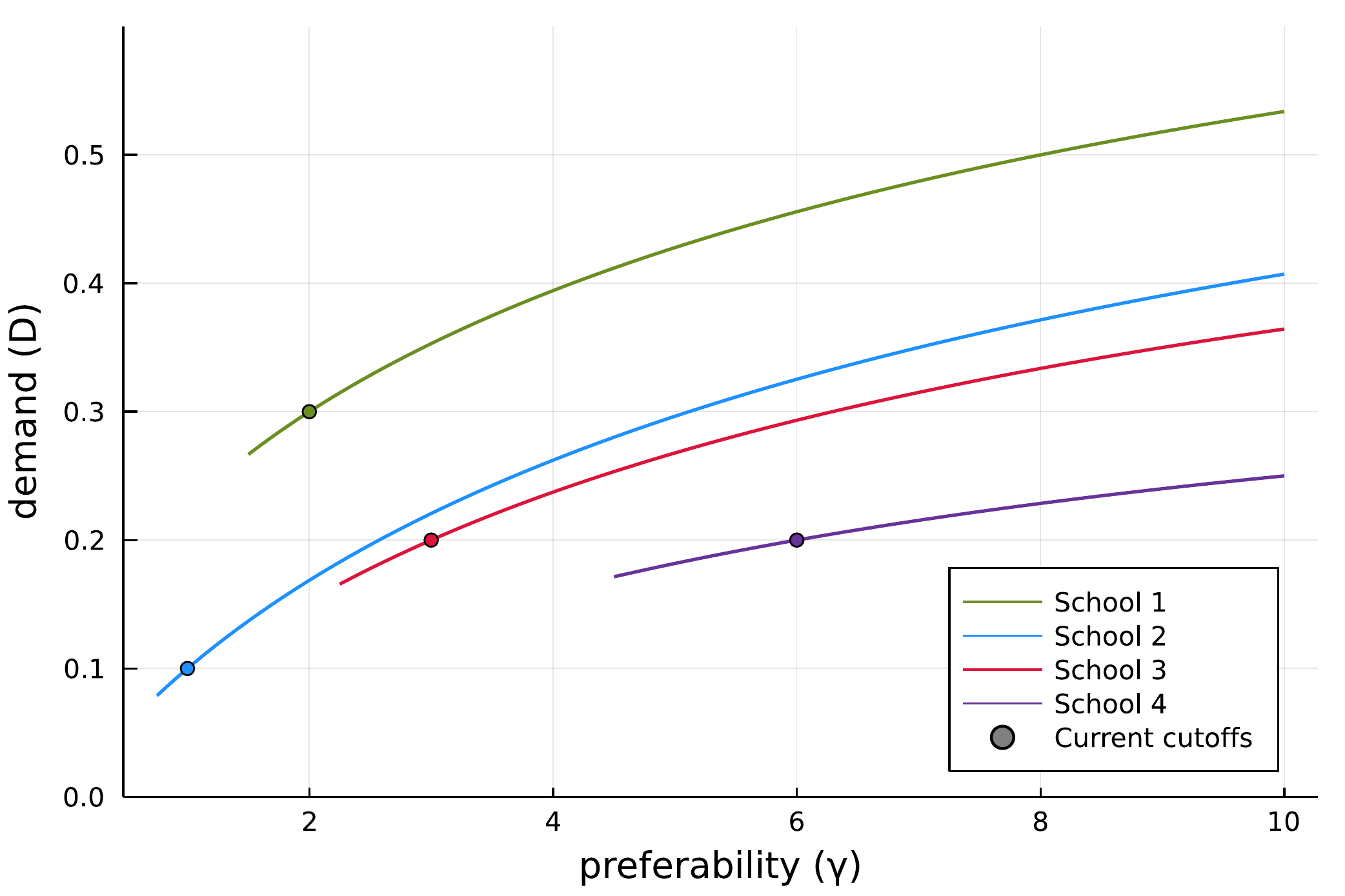}\end{center}
\captionsetup{singlelinecheck=off}
    \caption[.]{Quality effects in Pallet Town under a decentralized admissions process like that considered in Figure \ref{score-cutoff-choice}. Each line shows the change in each school’s demand when it changes its quality $\gamma_c$ while holding the cutoff vector and other schools’ quality fixed. To the extent that each school's goal is to increase its demand, the slope of the curve represents the strength the incentive to improve in quality. At the current quality vector, the Jacobian of the demand is
    \begin{equation*}
    \mathbf{J}_\gamma D = A = \frac{1}{360}
    \begin{bmatrix}
22 & -14 & -6&  -2\\
 -7   &29  &-3  &-1\\
 -9   &-9  &15  &-3\\
 -6   &-6&  -6  & 6
    \end{bmatrix}
    \end{equation*} 
    according to the expression provided in \S\ref{unconstrainedqualityeffects}. }
\label{vary-gamma-demand}
\end{figure}

\begin{figure}
\begin{center}\includegraphics[width=\linewidth, ]{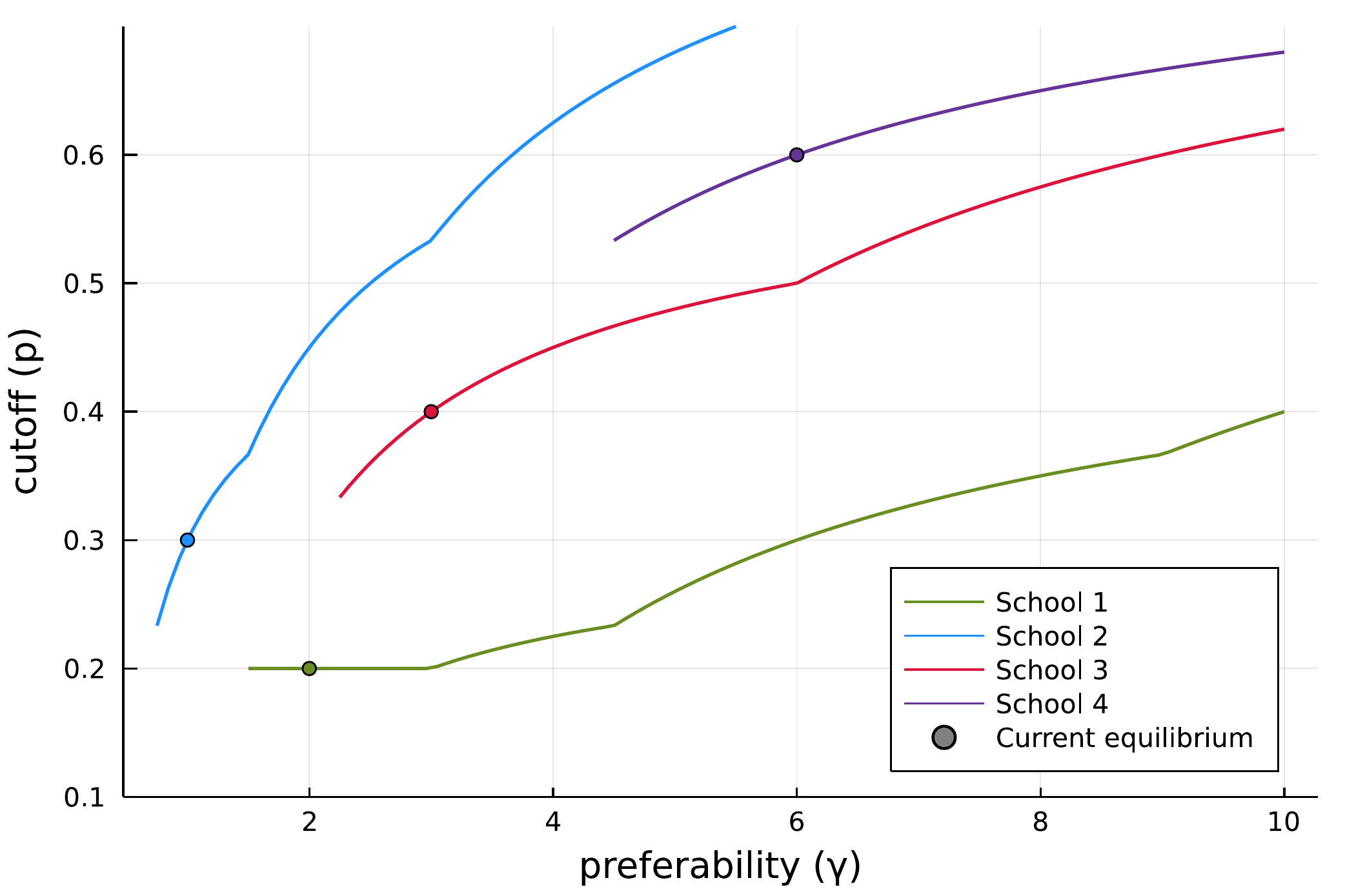}\end{center}
\captionsetup{singlelinecheck=off}
    \caption[.]{Quality effects in Pallet Town under a centralized admissions process such as the deferred acceptance process shown in Figure \ref{score-DA-placement}. Each line shows the change in each school’s equilibrium cutoff when it changes its quality $\gamma_c$ while holding other schools’ quality fixed. To the extent that each school’s objective is increase its cutoff, the slope of the curve represents the strength of the incentive to improve in quality. At the current quality vector, the Jacobian of the equilibrium cutoffs is
    \begin{equation*}
    \mathbf{J}_\gamma \hat p = \frac{1}{30}
    \begin{bmatrix}
      0 & & & \\
      -3 & 6 & & \\
      -2 & -2 & 2 & \\
      -1 & -1 & -1 & 1 
    \end{bmatrix}
    \end{equation*} 
according to the expression derived in \S\ref{qualityeffectsateq}. The elbows in the graph correspond to ties among the competitiveness ratios. 
    }
\label{vary-gamma-cutoff}
\end{figure}

Figures \ref{vary-gamma-demand} and \ref{vary-gamma-cutoff} visualize the incentive analysis under two different assignment paradigms. In Figure \ref{vary-gamma-demand}, the market is decentralized. Suppose that each school would like to achieve higher demand, and that no school is willing to lower cutoff beyond its current value. (For the sake of comparison with the figure that follows, these fixed cutoffs are chosen to equal the equilibrium cutoffs, although the decentralized market allows schools to exceed their capacity.) The four curves in the graph show how each school's demand would respond to a change in each school's $\gamma_c$-value. The slope of each school's quality--demand curve is given by the diagonal of the Jacobian given in equation \eqref{jac-gamma-demand-uncons}. All schools have a positive incentive to improve in quality, and school 2's incentive is the strongest.

Figure \ref{vary-gamma-cutoff} considers a market that is confined to equilibrium, which can be interpreted as a centralized market that uses a DA mechanism or as the equilibrium of a competitive market in which schools' capacities represent target class sizes or hard constraints. In this case, each school is already filled to capacity, so no school has any hope of increasing its demand; instead, suppose that schools hope to increase achieve a higher equilibrium cutoff by increasing their quality. The $y$-axis of the figure shows how each school's equilibrium cutoff responds to an increase in quality; the slope at the current parameters is given by the diagonal of the matrix given in equation \eqref{jac-gamma-p}.

Perhaps the most interesting feature of Figure \ref{vary-gamma-cutoff} is the flat slope at school 1. When the single-score admissions market is confined to equilibrium, the bulk of students who attend school 1 are those who could not obtain admissions anywhere else. A marginal improvement in school 1's quality pulls some students away from more desirable schools, but the \emph{minimum} score of students at school 1 (that is, its cutoff) does not increase until school 1's quality increases to the point that its competitiveness ratio exceeds that of school 2. The plot shows a distinctive set of ``elbows'' where these transitions occur. Visually, these elbows confirm that the market-clearing cutoffs are continuous in the market parameters.

\subsection{Inverse optimization of single-score admissions markets}
In this section, I consider the inverse optimization task, in which the demand of the market and the cutoff vector is provided and we attempt to compute the quality parameters. I provide an analytic solution, discussion its usefulness to admissions planners, and offer a proof-of-concept demonstration of the inverse optimization task using admissions data from 677 American universities. 

\subsubsection{The inverse optimization task and an analytic solution}
Given the demand $D$ and cutoff vector $p$, we must solve the following system for $\gamma$.
\begin{equation}
D_c = \sum_{d=c}^{|C|} 
\frac{\gamma_c}{ \sum_{i=1}^d \gamma_i} 
\left(p_{d+1} - p_{d}\right),
\quad \forall c \in C
 \label{solvemeforgamma}
 \end{equation}
Assume the schools are sorted in ascending order of cutoffs, and by homogeneity, let $\sum_{c \in C} \gamma_c \equiv 1$.

Consider the demand for $|C|$, the school with the highest index and therefore highest cutoff. Students who get into this school necessarily get into every school, so the outer sum of the system \eqref{solvemeforgamma} has only one term, and the equation becomes
\begin{align}D_{|C|} =
\frac{\gamma_{|C|}}{ \sum_{i=1}^{|C|} \gamma_i} 
\left(1 - p_{|C|}\right) \\
\implies \gamma_{|C|} = \frac{D_{|C|}}{1 - p_{|C|}}
\end{align}
Now suppose that we know $\gamma_{c+1}, \gamma_{c+2}, \dots, \gamma_{|C|}$. Then $\gamma_c$ can also be calculated from the observation that
\[\sum_{i=1}^d \gamma_i = 1 - \sum_{j=d+1}^{|C|} \gamma_j\]
where I take $\sum_{j=|C|+1}^{|C|} \gamma_j \equiv 0$.

Hence, the following recursive relation allows us to compute all the $\gamma_c$ values in reverse order, starting with $\gamma_{|C|}$ and moving down.\footnote{Because this expression requires repeated division by small numbers, it is numerically unstable when the number of schools is large or there are many schools whose cutoffs are close together or equal. Moreover, the error accumulates with each iteration. Thus, it is sometimes more effective to solve the system \eqref{solvemeforgamma} using a generic root finder; this spreads the numerical error out over all the schools.} 
\begin{align}
\gamma_{|C|} &= \frac{D_{|C|}}{1 - p_{|C|}} \\[1em]
\gamma_c &= \frac{D_c}{~~\mathlarger{\sum_{d=c}^{|C|} \frac{p_{d+1} - p_d}{1 - \sum_{j=d+1}^{|C|} \gamma_j} ~~}}, \quad \forall c \in \bigl\{ |C|-1, |C|-2, \dots, 1\bigr\} \label{gammarecursion}
\end{align}

\subsubsection{A demonstration using admissions data from American universities}
Here, I demonstrate the inverse optimization process using a public-domain admissions dataset from a large set of American universities \parencite[][]{collegeadmissionskaggle}. The results and discussion below should be taken only as a proof of concept, for four reasons: First, this model makes the unrealistic assumption that all colleges have the same preference order, which is derived from students' standardized test scores. Second, I did not attempt to account for the fact that many students do not bother applying to schools for which they are are overqualified; as a result, the cutoffs are systematic underestimates. Third, the dataset is not adequately sourced, and appears to mix data across years. I used data from the 2017 ACT and 2014 SAT examinations to derive school cutoffs \parencite[][]{ACTprofilerpt, SATpercentileranks}. Finally, I excluded from consideration schools for which test statistics were not listed, reducing the dataset from 1534 schools to 677. Thus, whereas the inverse optimization procedure provided above estimates schools' \emph{preferability} when given perfect knowledge of their \emph{selectivity}, in the present analysis, both quantities had to be estimated.

The dataset contains information regarding four test scores: the critical reading, math, and writing SAT subscores, and ACT composite scores. For each score, the dataset shows the 25th and 75th percentile of scores among students admitted to each school. By comparing these percentiles to percentile tables released by the testing agencies, I derived eight different estimates of each school's cutoff value. For example, at the University of Alabama, the 75th-percentile ACT composite score among admitted students is a 30 (out of 36). Relative to the population of ACT examinees, a student who earns a 30 is at the 90th percentile overall; hence, if the University of Alabama looks only at ACT composite scores, it follows that the minimum score among Alabama's admits is at the 60th percentile among all test-takers. In the occasional case where this calculation yielded a negative value, I cropped it to zero. Explicitly, letting $p_{\text{rel}}$ (here $0.75$) denote the percentile under consideration, and $p_{\text{abs}}$ (here $0.90$) denote the percentile score of a student with that score relative to the whole student population, the implied school cutoff is \[p_{\text{impl}} = \max\left\{0, 1 - \frac{1 - p_{\text{abs}}}{1- p_{\text{rel}}}\right\}\]

To compute a composite estimate of each school's cutoff, I first averaged the cutoffs implied by the ACT and SAT data separately. Then, I took the average of these two weighted by the percentage of applicants who submitted each test score (which is also included in the dataset) and treated this as the school's $p_c$. Schools missing data for any of the eight data points mentioned above were excluded from consideration.

To compute each school's demand, I divided the number of students enrolled at each school by the sum of the same, which is 752,987. Thus, this model assumes that every student in the market can get into at least one college, and that the dataset includes all the college-like options that students in this market would consider. The first assumption is mild, because there are several colleges in the dataset whose estimated cutoff is zero. The second assumption is much more restrictive, as there are many colleges that do not report test scores, not to mention international schools, that are not represented in the data. (In the tabular results and graphs, I report demand as a number of students instead of the proportion $D_c$ for legibility.)

\begin{figure}
\begin{center}\includegraphics[width=\linewidth, ]{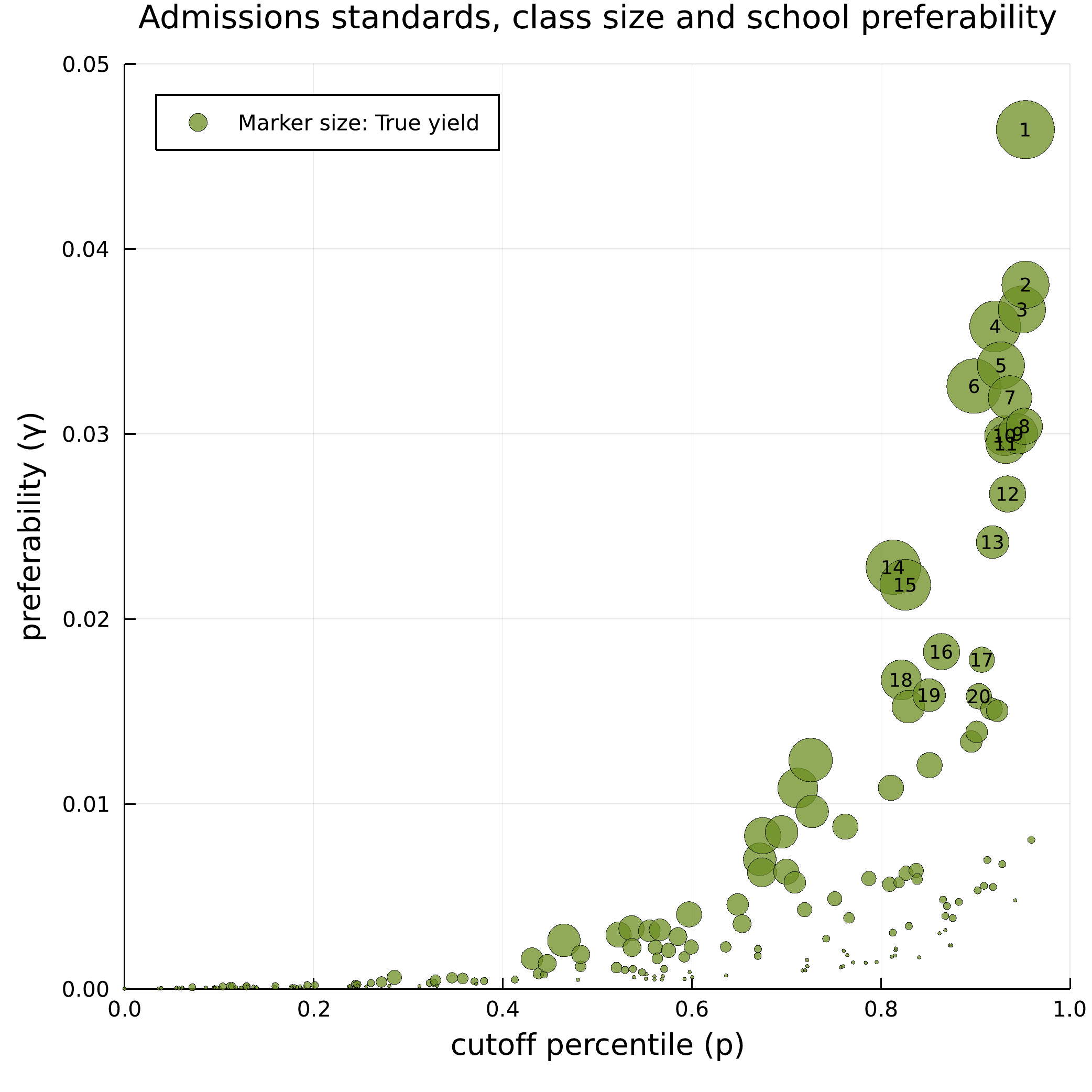}\end{center}
\captionsetup{singlelinecheck=off}
    \caption[.]{Inverse optimization procedure applied to a dataset of 677 American universities. School cutoffs were determined using a weighted average of cutoffs inferred from admitted students' SAT and ACT scores at the 25th and 75th percentiles. The marker size indicates the true yield, or percentage of qualified students who choose to attend each school. Details for the top twenty schools, as ranked by $\gamma_c$, appear in Table \ref{tab:US-inverse-optimization}.}
\label{US-cutoff-gamma}
\end{figure}

The results are shown in Figure \ref{US-cutoff-gamma} and Table \ref{tab:US-inverse-optimization}. The list of the schools with the highest preferability is a predictable list of prestigious universities. This outcome is remarkable, because the input data contains no explicit indicators of ``prestige'' (such as data on endowment size and employment outcomes), nor any survey data akin to the questionnaires that some newspapers send to college executives to generate their rankings. The model also does not require observations of individual student choices, as used to estimate MNL parameters under traditional statistical paradigms. The preferability parameters simply emerge organically from the selectivity of each school, each school's total demand, and the assumptions about the distribution of student talent. For comparison, Table \ref{tab:US-inverse-optimization} provides other metrics that might be used to assess school preferability such as the demand, cutoff, yield (proportion of admitted students who choose to attend), and \emph{true yield.} The last is a contrived term for the proportion of \emph{qualified} students, regardless of whether they applied, who chose to attend the school; this can be computed from the cutoff and demand. 

Again, due to the low quality of the underlying data, I caution against reading deep meaning into this model's estimates. However, it is worthwhile to take the results at face value for a moment to demonstrate the sort of descriptive analysis enabled by the inverse optimization procedure. 

First, consider the top-ranked schools, whose figures are shown in Table \ref{tab:US-inverse-optimization}. In this market, the top two, Harvard and Yale, appear equally selective (they have almost the same cutoff). However, Harvard achieves a higher preferability parameter because at the same admissions standards, it attracts a larger student body (1659 students, versus Yale's 1356). Conspiciously absent from the list is the school with the highest cutoff, California Institute of Technology ($p_c = 0.9590$, $\gamma_c = 0.0081$, rank 34). While Caltech is a little more selective than Harvard, its entering class size is much smaller, at 249 students. Since, in this model, the set of students admitted to Caltech is only a slight subset of those admitted to Harvard, Harvard must be about $1659 / 249 = 6.7$ times as preferable. Perhaps an admissions director at Caltech would argue that the school's small entering class size is a key element of its appeal, and this is certainly true for many students. However, in this model, Harvard could easily achieve a similar class size to Caltech at a much higher cutoff, eliciting the same conclusion.

 Because this model considers not only selectivity but also entering class size as an indication of market power, compared to conventional college rankings, it grants an elevated position to public flagships like the University of Michigan at Ann Arbor, which draw large entering classes while maintaining fairly high admissions standards. Although tuition price has not figured into any stage of the analysis, this example shows that the computed preferability parameters incorporate a notion of \emph{value,} whereas traditional college rankings arrange schools as if their tuition prices are the same.

\begin{figure}
\begin{center}\includegraphics[width=\linewidth, ]{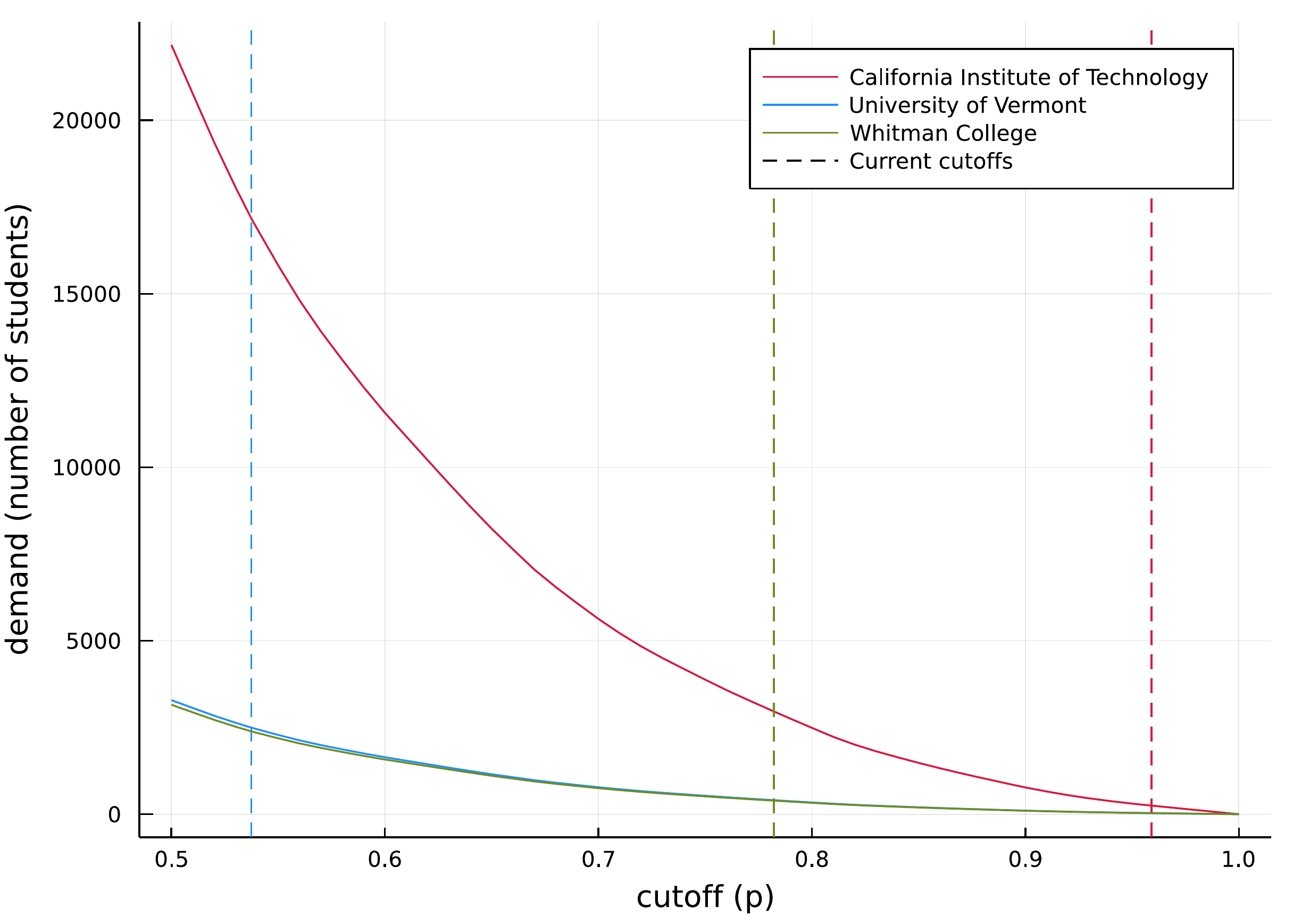}\end{center}
\captionsetup{singlelinecheck=off}
    \caption[.]{Three demand curves derived via the inverse optimization process. The University of Vermont and Whitman College have similar preferability parameters, and thus similar demand curves. However, each school has chosen a different selectivity threshold, reflecting its distinct admissions priorities. A detailed view of Caltech's demand curve appears in Figure \ref{caltech-demand-curve}.}
\label{three-demand-curves}
\end{figure}

It is worth considering schools in other parts of the preferability distribution. While conventional wisdom posits small liberal-arts colleges and large public universities as incommensurable, administrators at both types of school share a common goal of recruiting an entering class that is both ``large'' relative to the physical size of the campus and ``highly qualified'' relative to competitor schools. $\gamma_c$ offers us a way to compare the effectiveness of two schools' marketing efforts even when their recruitment strategies diverge. For example, the University of Vermont and Whitman College (a liberal-arts college in eastern Washington) rank 106th and 107th, with preferability parameters $1.079 \times 10^{-3}$ and $1.044 \times 10^{-3}$, respectively. Vermont has a large class size and a middling cutoff, while Whitman has a small class size and a cutoff close to that of UM Ann Arbor. Looking only at conventional statistics, it is hard to predict the decision of a student choosing between the two schools, but comparing $\gamma$-values (which are, by definition, choice probability weights) reveals that in this case it is a nearly even coin flip. Indeed, the two school's demand curves (shown in Figure \ref{three-demand-curves}) are all but identical. 

The inverse optimization task makes no equilibrium assumption, and indeed invokes no notion of capacity or target class size. It simply reports the status of the market with respect to the current allocation of students. Thus, a possible application of this model is for an admissions director to use in modeling her school's demand curve. Figure \ref{three-demand-curves} shows the predicted demand curves for Vermont, Whitman, and Caltech. In the future, suppose that Caltech decides that a larger cohort of 350 students better suits its goals. By how much should it relax its admissions standards in order to achieve this class size? One way to answer this question is to assume that Caltech’s true yield remains approximately fixed. Then, Caltech should try to become $\frac{350}{249}$ as selective, by updating its cutoff to $1 - \frac{350}{249}(1 - 0.9590) = 0.9424$. A simple calculation shows that this way of estimating the demand curve is equivalent to fitting a linear model to the observed demand $(p_c, D_c) = (.9590, 249)$ and the implicit $x$-intercept $(1, 0)$. 

However, when using a linear model, the recommended cutoff associated with the higher target demand will be a slight underestimate, because Caltech’s true yield also varies as a function of $p$. Under the recommended cutoff, students admitted with scores of (say) $0.95$ do \emph{not} qualify for Harvard and Yale, so Caltech will not have to compete as fiercely to recruit them as it does to recruit its current enrollees. The model presented here accounts for this change in the consideration set of marginal students, and thus calls for a more modest reduction in Caltech’s cutoff, to the value of $0.9437$.

\begin{figure}
\begin{center}\includegraphics[width=\linewidth, ]{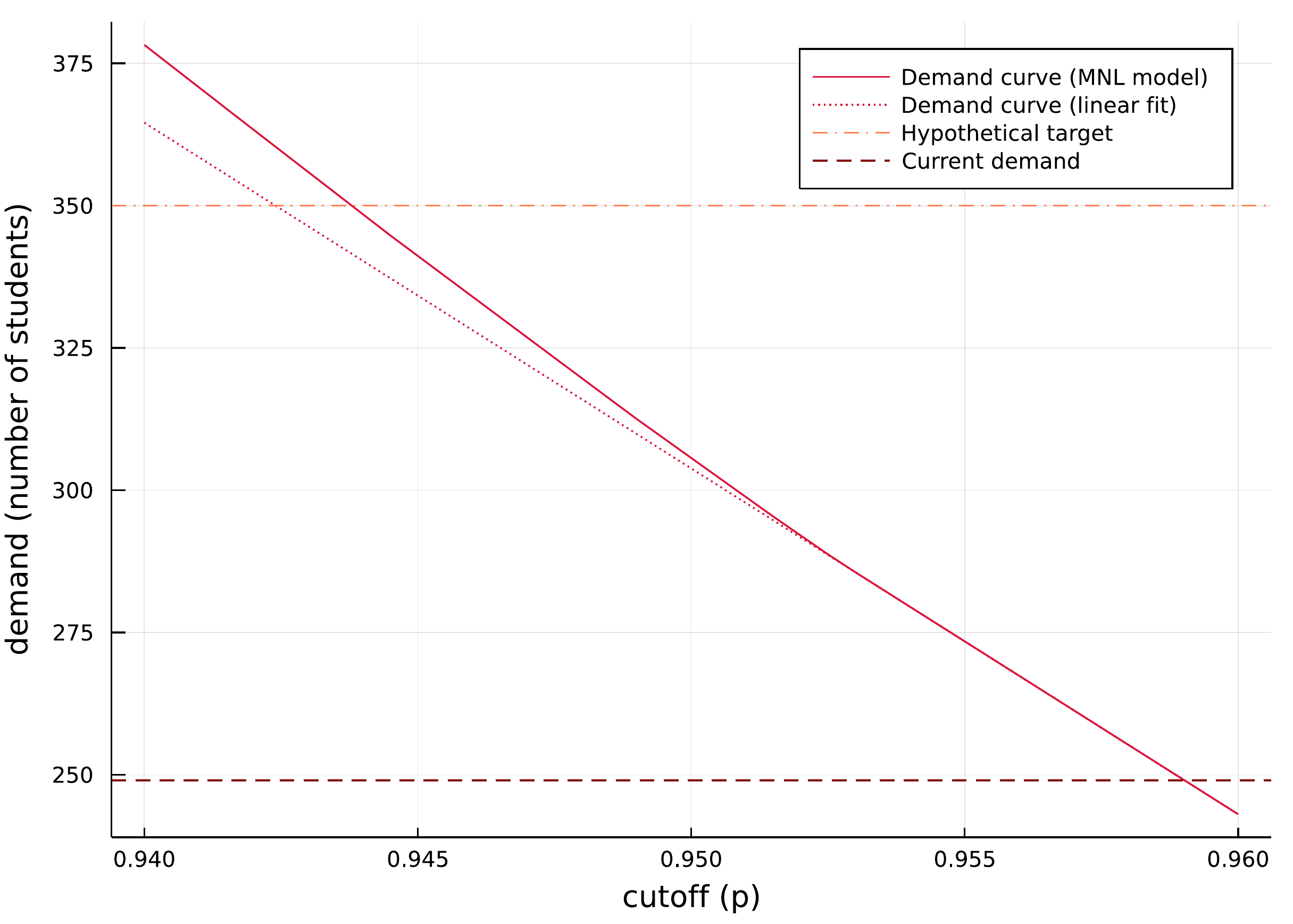}\end{center}
\captionsetup{singlelinecheck=off}
    \caption[.]{Detailed view of Caltech's demand curve near its current cutoff. Assuming preferability and other schools' cutoffs remain fixed, if Caltech wishes to increase its class size to 350, it should update its cutoff to the value indicated by the intersection of the demand curve and the horizontal coral line. A linear model prescribes decreasing the cutoff to $p_c = 0.9424$, whereas the model provided here accounts for the fact that there is less competition for marginal students at the lower cutoff and prescribes a more conservative decrease to $0.9437$. }
\label{caltech-demand-curve}
\end{figure}

Figure \ref{caltech-demand-curve} presents a detailed view of Caltech’s demand curve as predicted by this model alongside the linear model. The curve has a slightly bowed shape, which confirms the intuitive argument for a gentler relaxation of admissions standards in order to achieve lower target enrollment. In fact, every school’s demand curve in this model is piecewise linear convex, meaning that linear regression will always underestimate the demand at cutoffs far from those used to fit the line.\footnote{If the linear model is constrained to contain the point $(1, 0)$, then it forms a chord of the convex curve, and it will \emph{overestimate} the demand at $p_c$-values \emph{higher} than those used to fit the model. On the other hand, if this constraint is dropped and the model is fitted to two or more local observations, then it will \emph{underestimate} the demand at both higher and lower cutoffs.} Of course, a more accurate regression could be constructed by using demand observations from multiple years and including a quadratic term. But even then, if the observations used to fit the curve remain in the same ``piece'' of the piecewise linear demand function, then the expected regression curve will be a straight line.

This analysis has not accounted for the hypothesis that Caltech's appeal depends on a small class size, in which case looser admissions standards also reduce the school's preferability, necessitating a lower cutoff after all. Thus, in practical decisionmaking a model like that presented here is unlikely to be competitive with colleges' in-house models, which incorporate specific observations of students who applied to the school and chose to attend another. However, an advantage of the current model is that it produces an informative approximation of the demand curve without using students' personal data. Thus, a school like Caltech can use it to model the demand curves of its \emph{competitors} schools, enabling a more sophisticated recruitment strategy. 

\begin{table}[]
\begin{tabular}{r|l|r|l|l|l|l}
\multicolumn{1}{l|}{\textbf{Rank}} & \textbf{University}                   & \multicolumn{1}{l|}{\textbf{Demand}} & \textbf{\begin{tabular}[c]{@{}l@{}}Cutoff\\ ($p_c$)\end{tabular}} & \textbf{Yield} & \textbf{True yield} & \textbf{\begin{tabular}[c]{@{}l@{}}Preferability\\ ($\gamma_c$)\end{tabular}} \\ \hline
1                                  & Harvard U                             & 1659                                 & 0.9526                  & 0.81           & 0.0465              & 0.0465                              \\
2                                  & Yale U                                & 1356                                 & 0.9527                  & 0.66           & 0.0381              & 0.0381                              \\
3                                  & U of Chicago                          & 1426                                 & 0.949                   & 0.53           & 0.0371              & 0.0367                              \\
4                                  & U of Pennsylvania                     & 2421                                 & 0.9207                  & 0.63           & 0.0405              & 0.0358                              \\
5                                  & Northwestern U                        & 2037                                 & 0.9268                  & 0.41           & 0.0369              & 0.0337                              \\
6                                  & Cornell U                             & 3223                                 & 0.8984                  & 0.52           & 0.0421              & 0.0326                              \\
7                                  & Washington U in St. Louis              & 1610                                 & 0.9364                  & 0.34           & 0.0336              & 0.032                               \\
8                                  & Mass. Institute of Technology & 1115                                 & 0.9515                  & 0.72           & 0.0305              & 0.0304                              \\
9                                  & Princeton U                           & 1285                                 & 0.9445                  & 0.65           & 0.0308              & 0.03                                \\
10                                 & Stanford U                            & 1677                                 & 0.9307                  & 0.76           & 0.0321              & 0.0299                              \\
11                                 & Vanderbilt U                          & 1613                                 & 0.932                   & 0.41           & 0.0315              & 0.0295                              \\
12                                 & Columbia U    & 1415                                 & 0.9338                  & 0.6            & 0.0284              & 0.0268                              \\
13                                 & Duke U                                & 1714                                 & 0.918                   & 0.42           & 0.0278              & 0.0242                              \\
14                                 & U of Michigan--Ann Arbor               & 6200                                 & 0.813                   & 0.4            & 0.044               & 0.0228                              \\
15                                 & New York U                            & 5207                                 & 0.8256                  & 0.35           & 0.0397              & 0.0218                              \\
16                                 & Northeastern U                        & 2891                                 & 0.864                   & 0.19           & 0.0282              & 0.0182                              \\
17                                 & Brown U                               & 1543                                 & 0.9065                  & 0.58           & 0.0219              & 0.0178                              \\
18                                 & U of California--Berkeley              & 4162                                 & 0.8214                  & 0.37           & 0.0309              & 0.0167                              \\
19                                 & U of Southern California              & 2922                                 & 0.8509                  & 0.31           & 0.026               & 0.0159                              \\
20                                 & Carnegie Mellon U                     & 1442                                 & 0.9035                  & 0.3            & 0.0198              & 0.0158                             
\end{tabular}
\caption{\label{tab:US-inverse-optimization}
The top twenty schools by preferability $\gamma_c$, as determined by applying the inverse optimization process to a dataset of 677 American universities. Each school's demand is given as the number of students in the entering class; to compute $D_c$, divide by the total number of students, 752,987. The school's yield is as reported by the admissions office, while the true yield, which represents the proportion of qualified students who chose to attend the school, was computed by comparing the size of the entering class to the estimated cutoff $p_c$.}
\end{table}

The code used to produce these results, as well as more details regarding how the cutoffs were estimated from available SAT and ACT data, is available on GitHub \parencite[][]{studentprefsrevopt}.

\subsubsection{The informational quality of $\gamma$, and the bias in (true) yield}
The question of how to measure aggregate college preferability is not a simple one, and newspaper university rankings attract regular controversy for their imprecise methodology \parencite[][]{intlrankingsandconflicts}. Typical ranking metrics draw from a combination of survey data and performance measures of a university's quality such as its research output or data on alumni salaries. However, conducting accurate surveys is costly and difficult, and while performance measures may correlate with college preferability, they offer a normative indication of which colleges ``should'' be popular without accounting for the decisions students actually make---decisions that may depend less on hard facts than on intangible notions of fitness.

The traditional measure college administrators have used to quantify how preferable their school is relative to others in the market has been the yield. As discussed above, colleges' \emph{true} yield (or another yield metric that corrects for applicant behavior) can be a useful tool in modeling the demand curve. However, as a measure of comparing the preferability of two different colleges, true yield systematically overrates lenient schools, which face less competition for students. For example, consider a market with only two schools, Lunar College ($p_1 = 0, D_1 =  \frac{100}{101}$) and Antarctic University ($p_2 = \frac{99}{100}, D_2 = \frac{1}{101}$). At both schools, the true yield is $D_c / (1 - p_c) = \frac{100}{101}$, a value which exaggerates the preferability of Lunar College by failing to account for the fact that the majority of its admits, and the vast majority of its enrollees, had no other option. The proportion of students who chose Lunar College \emph{over} Antarctic University is given by the former's preferability, $\gamma_1 = \frac{100/101 - 99/100}{1/100} = \frac{1}{101}$, while the preferability of Antartic University is $\gamma_2 = \frac{100}{101}$. This example, along with the demonstration above, suggests that insofar as $\gamma$ can be estimated accurately, it provides unbiased information about school preferability in a well-differentiated admissions market.

\section{Discussion}
This article has considered the characterization of nonatomic admissions markets, and attempted to incorporate the best elements of the mechanism-design paradigm that treats stable assignment as a normative goal alongside those of regression models that parameterize colleges' demand in a decentralized context. Under the assumption that each school's capacity in the school-choice problem can be interpreted as its target demand, stable matchings produced by DA mechanisms coincide with competitive equilbria of decentralized admissions markets. While this interpretation readily follows from prior results, I have also argued that each iteration of DA mechanisms can be characterized by the price vector, and thus that DA is a special form of t\^{a}tonnement. Thus, given access to an oracle that computes the demand for each school at a given cutoff vector, equilibrium cutoffs can be efficiently computed using a generic t\^{a}tonnement algorithm.

One interpretation of the relationship between discrete admissions markets (with a finite collection of students and integral school capacities) and nonatomic admissions markets (with a distribution over the set of student types and fractional school capacities) regards instances of the former as samples from the latter. Hence, if the parameters of a given economy are known, then computing equilibria and comparative statics with respect to the nonatomic formulation offers a picture of the expected behavior of the market that is insensitive to discretization error. Unfortunately, as shown in this paper's first section, the endogenous complexity of the space of student types makes nonatomic markets difficult to work with directly. Thus, constructing the oracle mentioned in the previous paragraph is usually difficult, and there remains no general solution to the problem of finding equilibrium cutoffs in an arbitrary nonatomic market.

Nonetheless, by making simplifying assumptions on schools' scoring practices and choosing a suitable parametric choice model, we can sometimes compute the demand efficiently. In the example considered here, students choose schools according to a multinomial logit function and schools share a common scoring procedure. Then an invertible, piecewise linear expression relates the market parameters $\gamma$, $D$, and $p$. Though this model is rather simplistic, solving for the preference parameters $\gamma$ using real data from a set of American colleges produced an intuitive ordering of top universities. The model also allows the analyst to chart each school's demand curve without consulting granular data on individual students' enrollment decisions or assumptions about either set of players' utility functions. 

\subsection{Future directions}
Natural extensions of the computational model considered here include the mixed MNL model and mixed scoring vectors. Both of these modifications would allow for differentiation based on students' academic interests; for example, students interested in the performing arts systematically promote conservatories in their preference orders, and these schools in turn will score applicants according to different criteria. However, any departure from the single scoring model invites rapid multiplication in the number of possible consideration sets, in principle because it becomes impossible to construct a total ordering of the space of students. Further upstream sits the question of how to evaluate student ability in the first place. Modern standardized examinations used to assess school performance are weighted by individual and school characteristics, and the scoring formula treats the high-dimensional concatenation of student responses, rather than simply the number of correct answers, as the explanatory variable of ability \parencite[][]{measurementofstudentability}. These devices complicate the analyst's task.

A more robust account of the notion of equilibrium in decentralized admissions markets is also needed. This paper's definition of equilibrium was chosen because it coincides with stable assignments, but in principle, schools can have arbitrary utility functions, and the equilibrium in general nonatomic admissions markets will seldom agree with the stable assignment produced by a centralized mechanism. Furthermore, real schools' preferences may be cardinal rather than ordinal, and are sensitive to the composition of the entering class along with its summary statistics, as pointed out by Roth \parencite*{collegeadmissionsisnotmarriage}. Thus, the results given in this paper cannot fully predict the relative efficiency and fairness of decentralized admissions procedures. 

However, in the real world, putatively decentralized admissions procedures often incorporate regulatory elements that push the market toward stable assignment. For example, the college admissions procedure in South Korea can be viewed as a distributed approximation of school-proposing DA. In this model, the government places a firm limit, called an admissions quota, $q_c$ on the number of students who can attend each university. At the beginning of the admissions cycle, colleges are given the profiles of all the students interested in attending. Each college makes admissions offers over the course of several rounds, beginning with the highest-qualified students, and at each round a subset of the admitted students tentatively commits to attending one of the colleges that admitted them. This process continuous for three rounds, at which point most colleges have either filled their capacity, or bottomed out their pool of applicants. If students were allowed to apply to every college and there were infinite rounds, this would be equivalent to school-proposing DA. However, students are in fact allowed to apply to only six colleges, and colleges instate different evaluation criteria for students who rank the school with high priority. It would be interesting to quantify the welfare cost of these regulations relative to a true stable assignment procedure. The MNL model with single scoring could be a useful analytic tool, because admissions decisions in Korea are based heavily on standardized test scores, and the market is well differentiated vertically.

This article has taken a microeconomic view of admissions markets that views the market as a closed system in which the set of schools and the distribution of student preferences are frozen. Thus, the results offered here are incommensurate with macroeconomic studies in the so-called Tiebout framework that model the sorting behavior of families who move in and out of admissions markets according to how much they value public education overall \parencite[][]{apuretheoryoflocalexpenditures, equilibriumandlocalredistribution}. Because this area of the literature includes comparisons across districts that use different assignment mechanisms, it would be worthwhile to examine the relationship between sorting behavior and choice mechanisms, whereas previous regression studies have treated sorting behavior as a dependent variable that must be controlled for \parencite[][]{doescompetitionamongpublicschools}.

\pagebreak
\section{References}
\printbibliography[heading=none]

\end{document}